\documentclass[twoside,onecollarge]{article}

\usepackage[applemac]{inputenc}
\usepackage[dvips]{graphicx}
\usepackage{graphicx}
\usepackage[english]{babel}
\usepackage{amsfonts}
\usepackage[reqno]{amsmath}
\usepackage{epsfig}
\usepackage{float}
\usepackage{bbm}

\makeindex

\newtheorem{theorem}{Theorem}

\newcommand{\F}{\mathbb{F}}

\newcommand{\PP}{\mathbb{P}}

\newcommand{\R}{\mathbb{R}}

\newcommand{\Z}{\mathbb{Z}}
\newcommand{\E}{\mathbb{E}}

\newcommand{\boF}{\mathcal{F}}

\newcommand{\boL}{\mathcal{L}}

\newcommand{\bfE}{\mathbf{E}}
\newcommand{\bfF}{\mathbf{F}}

\newcommand{\rnf}{\renewcommand{\thefootnote}{\arabic{footnote}}}
\newcommand{\thankyou}[2]{\stepcounter{footnote}\footnotetext[#1]{#2}}

\usepackage{color}
\definecolor{lightgray}{gray}{0.85}
\newcommand{\Xbox}[2]{
\fcolorbox{lightgray}{lightgray}{\setlength{\fboxsep}{5pt}
\begin{minipage}{\textwidth}
\mbox{}
{\color{black}
#2
}
\end{minipage}
}
}

\title{High-frequency market-making with inventory constraints and directional bets}

\author{\rnf Pietro FODRA
              \footnotemark[1]
\and
\rnf Mauricio LABADIE
              \footnotemark[2]
}

\date{\today}

\begin{document}
\DeclareGraphicsExtensions{.pdf,.gif,.jpg} \maketitle

\thankyou{1}{\,EXQIM (EXclusive Quantitative Investment Management). 24 Rue de Caumartin 75009 Paris (France) and Universit\'e Paris-Diderot. 4 Place Jussieu 75005 Paris (France). E-mail: pietro.fodra@exqim.com}
\thankyou{2}{\,EXQIM. Corresponding author. E-mail: mauricio.labadie@exqim.com, mauricio.labadie@gmail.com}

%\thankyou{2}{\,Cheuvreux - Cr\'edit Agricole. Email address:  clehalle@cheuvreux.com}

%\thankyou{3}{\,Departamento de Matem\'aticas y Mec\'anica, IIMAS-UNAM. Apartado Postal 20-726,
%C.P. 01000 M\'exico D.F. (Mexico)\\
%E-mail address: plaza@mym.iimas.unam.mx}

%\thankyou{3}{\,Universit\'e Aix-Marseille III, LATP, Facult\'e de Sciences et Techniques. Avenue Escadrille Normandie-Niemen, 13397 Marseille Cedex 20 (France)\\
%E-mail address: francois.hamel@univ-cezanne.fr}

\maketitle

\begin{abstract}

In this paper we extend the market-making models with inventory constraints of Avellaneda and Stoikov (\emph{High-frequency trading in a limit-order book}, Quantitative Finance Vol.8 No.3 2008) and  Gu\'eant, Lehalle and Fernandez-Tapia (\emph{Dealing with inventory risk}, Preprint 2011) to the case of a rather general class of mid-price processes, under either exponential or linear PNL utility functions, and we add an inventory-risk-aversion parameter that penalises the marker-maker if she finishes her day with a non-zero inventory. This general, non-martingale framework allows a market-maker to make directional bets on market trends whilst keeping under control her inventory risk. In order to achieve this, the marker-maker places non-symmetric limit orders that favour market orders to hit her bid (resp. ask) quotes if she expects that prices will go up (resp. down).\

With this inventory-risk-aversion parameter, the market-maker has not only direct control on her inventory risk but she also has indirect control on the moments of her PNL distribution. Therefore, this parameter can be seen as a fine-tuning of the marker-maker's risk-reward profile.\

In the case of a mean-reverting mid-price, we show numerically that the inventory-risk-aversion parameter gives the market-maker enough room to tailor her risk-reward profile, depending on her risk budgets in inventory and PNL distribution (especially variance, skewness, kurtosis and VaR). For example, when compared to the martingale benchmark, a market can choose to either increase her average PNL by more than 15\% and carry a huge risk, on inventory and PNL, or either give up 5\% of her benchmark PNL to increase her control on inventory and PNL, as well as increasing her Sharpe ratio by a factor bigger than 2. 

\end{abstract}

\bigskip

\noindent {\bf Keywords}: Quantitative Finance, high-frequency trading, market-making, limit-order book, inventory risk, optimisation, stochastic control, Hamilton-Jacobi-Bellman, PNL distribution.

\maketitle

\section{Introduction}

\subsection*{Market-makers}

A market-maker is a trader who buys and sells assets in a stock exchange. The difference with any other market agent is that the market-maker is bound to make firm quotes: once she shows a buying/selling quantity at a certain price, she is engaged to trade under those conditions. As liquidity provider, a market-maker receives a compensation: she buys at a lower price (bid) and sells at a higher price (ask). This difference is called the \emph{spread}.\\

A market maker is exposed to two main risks, \emph{adverse selection} and \emph{inventory risk}. \emph{Adverse selection} means that if the market-maker sells (resp. buys) an asset it is not necessarily good news, it could mean that her ask (resp. bid) price is lower (resp. higher) than it should on the current market conditions. \emph{Inventory risk} comes into play by inbalances in the arrival of buying and selling orders: since the market-maker quotes both bid and ask prices, her net position depends on which quotes are executed and in which quantities. A market maker uses the spread to both control her inventory and compensate herself from adverse selection. In a nutshell, a market-maker loses money against informed traders, but she covers that loss by making noise (i.e. un-informed) traders pay the spread on each transaction.\\

In order to create a market-making strategy, we need to consider three factors: \emph{price}, \emph{spread} and \emph{inventory}. The \emph{price} is often the mid-price, i.e. the average between the current ask and bid prices of the market. The \emph{spread} of the market-maker is her only control on her PNL and \emph{inventory} throughout the trading day. It is true that the market maker can affect the mid-price by improving the current ask and bid market quotes, but since that would normally trigger a market order that consumes the offer, in a first approach we can consider that the mid price cannot be affected by the quotes of the market maker.

\subsection*{Main features of the present article}

In this article we extend the current stochastic-control models of market-making, in particular those of Avellaneda and Stoikov \cite{ave} and Lehalle \emph{et al} \cite{lehalle}.\\

We managed to find closed-form solutions for the optimal ask/bid quotes of a market-making for mid-price dynamics that are not necessarily martingale, which can be interpreted of directional bets on price trends. Although our approach is based on optimal stochastic-control and nonlinear-PDE techniques, the philosophy is very simple: given the utility function we choose a possible form of the solution of the nonlinear PDE equation (i.e. we make an \emph{ansatz}); we plug this \emph{ansatz} into the equation and compute the (implicit) optimal controls; we then plug the (implicit)controls and separate the equation into several simpler ones, normally linear; we then compute explicitly the controls and the solution for the equation.\\

Sometimes it is impossible to find explicitly the controls for the solution, but for a sub-solution it is always possible (at least for the class of processes and utility functions we are dealing with). In that framework, the controls do not optimise the utility function but give a lower bound of the potential losses. It is worth to mention that, for practitioners, it is better to have explicit controls that minimise potential losses (a sub-solution) than highly implicit, numerically-intensive controls that optimise the PNL.\\

Our approach can be applied to any utility functions, not only exponential as in Avellaneda and Stoikov \cite{ave} and Lehalle \emph{et al} \cite{lehalle}, provided it is explicit enough to admit an \emph{ansatz}. This gives a lot of flexibility to the market-maker for the choice of her risk-reward profile. Of course, selecting a mid-price dynamic determines the class of utility functions that we can choose from because the utility function has to be finite (a.s.). However, with the most recurrent mid-price models such as martingales, Brownian motion with drift, Ornstein-Uhlenbeck (resp. Black-Scholes) a linear (resp. exponential) utility function is finite.\\

We add a new parameter, which models the inventory-risk aversion of the market-maker. As it will be shown in the numerical simulations, this parameter guarantees that the trading algorithm will end the day with a flat inventory, which is the goal of a market-maker. Moreover, it also allows the market maker to control its directional bets via exposure to price movements intraday.\\

We show that the inventory-risk-aversion parameter not only exerces direct control on the inventory risk directly but it also has some indirect control on the risk in the PNL distribution of the market-maker (i.e. on the first four moments namely mean, variance, skewness and kurtosis). Moreover, this relation can be also inverted: the parameter of the exponential utility function has direct control on the PNL distribution and indirectly controls the inventory risk. This can be interpreted as a high risk - high reward scenario: big exposure to extreme events, either via fat tails or directional bets, improves the average PNL.

\subsection*{Organisation of the study}

The goal of this study is to find the optimal ask and bid quotes for a high-frequency market-maker that, under the framework of a directional bet on the market trend, simultaneously maximise her PNL and minimise her inventory risk. The main inspiration is the paper of Avellaneda and Stoikov \cite{ave}, who found via stochastic control the optimal bid and ask quotes for a high-frequency market-maker. The second inspiration comes from Lehalle \emph{et al} \cite{lehalle}, who formalised the findings of Avellaneda and Stoikov \cite{ave}. It is also worth to mention that the Hamilton-Jacobi-Bellman framework we use was originally set by Ho and Stoll \cite{ho}, but in neither of these three articles the effect of a directional market bet on the PNL distribution was considered.\\

In Chapter 2 we set the framework under which we will be working, which is stochastic control and Hamilton-Jacobi-Bellman equations. In Section 3 we completely solve the optimal control problem for a linear utility function and an arbitrary Markov process by finding explicitly the unique solution. In Section 4 we perturbate the linear utility function via a quadratic inventory penalty. By using first-order approximations on the arrival of orders to the limit-order book, we find explicitly the (approximate) solutions and their corresponding optimal controls, which are perturbations of the closed-form solution and controls we found in Section 3. In Section 5 we extend the results of Avellaneda and Stoikov \cite{ave} and Lehalle \emph{et al} \cite{lehalle} for exponential utility functions. We show that their approach can be used to more general process than arithmetic Brownian motion and with inventory-risk aversion. In Section 6 we perform numerical simulations to show how directional bets on the market trend affect the market-making strategies. We also assess the effect of the inventory-risk-aversion parameter on the PNL distribution and the inventory risk and show that this parameter controls directly the inventory risk and indirectly the PNL distribution.

\section{Stochastic control framework}

\subsection*{Setting of the problem}

We suppose that the mid-price process $S(t)$ follows an It\^o diffusion, i.e.
\begin{equation}\label{eq-pricedyn}
dS(t) = b(t,S(t))dt + \sigma(t,S(t))dW(t)
\end{equation}
where $W(t)$ is a standard Brownian motion in a filtered probability space $\left(\Omega,\boF,\F=(\boF_t)_{0\le t\le T},\PP\right)$.\\

A market-maker can control her ask and bid quotes, which we denote $p^+(t)$ and $p^-(t)$ respectively. Instead of working with the market-maker's prices we will rather work with the market-maker's spreads, i.e
\[
\delta^+(t):=p^+(t)-S(t),\qquad \delta^-(t):=S(t)-p^-(t)\,.
\]
The market-maker's spreads are assumed as two predictable processes. Under these variables, the market-maker's bid-ask spread is thus $\delta^++\delta^-\ge0$ (see Figure 1).

\begin{center}
\includegraphics[width=3.2in]{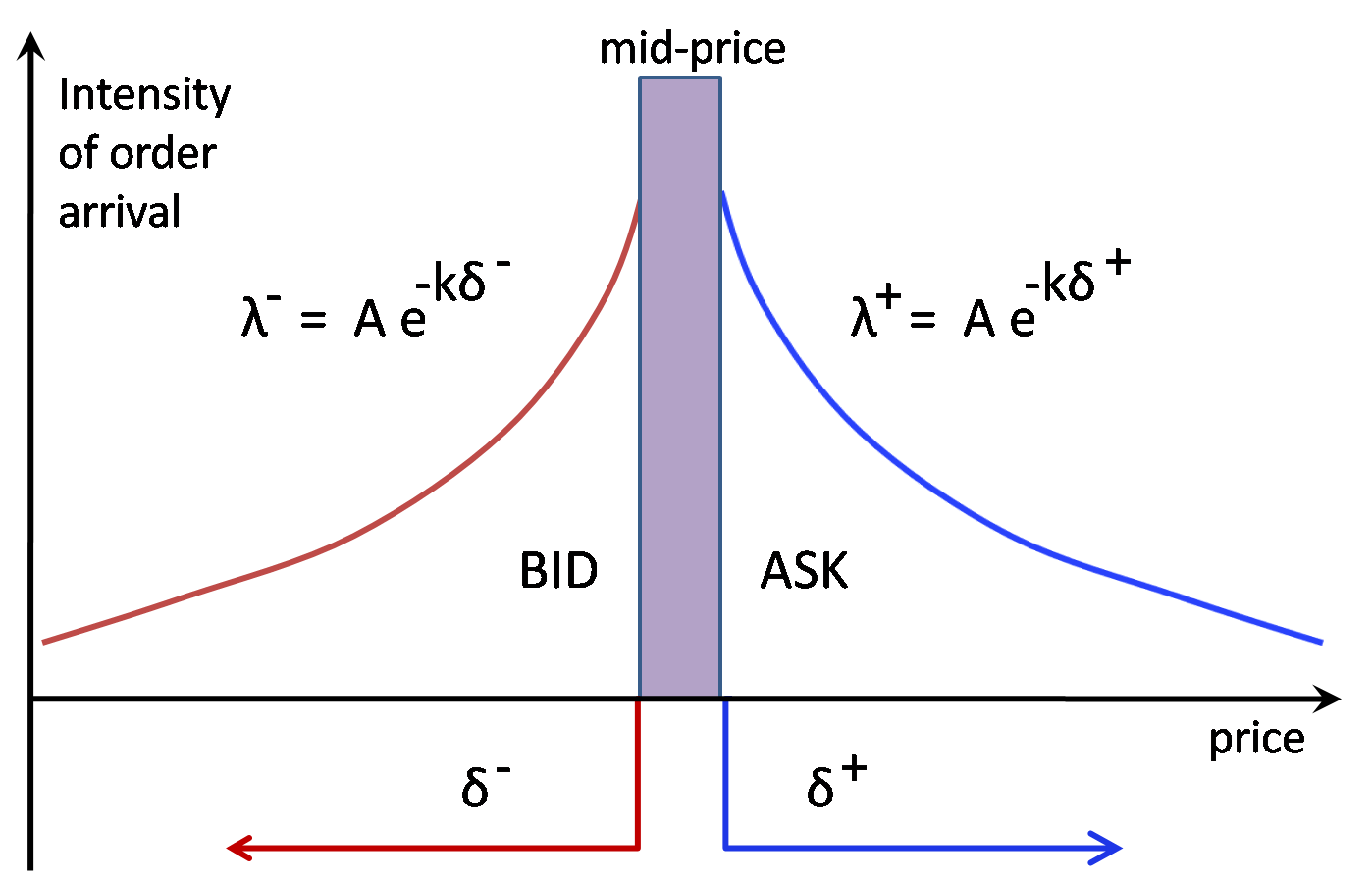}\\
Figure 1. Description of the intensity $\lambda^\pm$ of limit orders (LOs) as a function of the distances $(\delta^+,\delta^-)$. As usual, $\delta^\pm\mapsto\lambda^\pm(\delta^\pm)$ is decreasing: the closer we are to the mid-price, the more likely our LOs are executed. Here we explicited the exponential decay we will use in our model.
\end{center}

Strictly speaking, we should consider that $\delta^+\ge0$ and $\delta^-\ge0$. However, these constraints render the optimisation problem very hard to solve explicitly due to boundary effects at $\delta^\pm=0$. In consequence, and since our goal is to have explicitly the values of the optimal distances $(\delta^+_\ast,\delta^-_\ast)$ of the optimal market-maker quotes $(p^+_\ast,p-_\ast)$, we will assume in our analysis that $\delta^\pm\in\R$. However, in our numerical simulations we interpret $\delta^\pm\le0$ as a market order (see Figure 2).

\begin{center}
\includegraphics[width=3.2in]{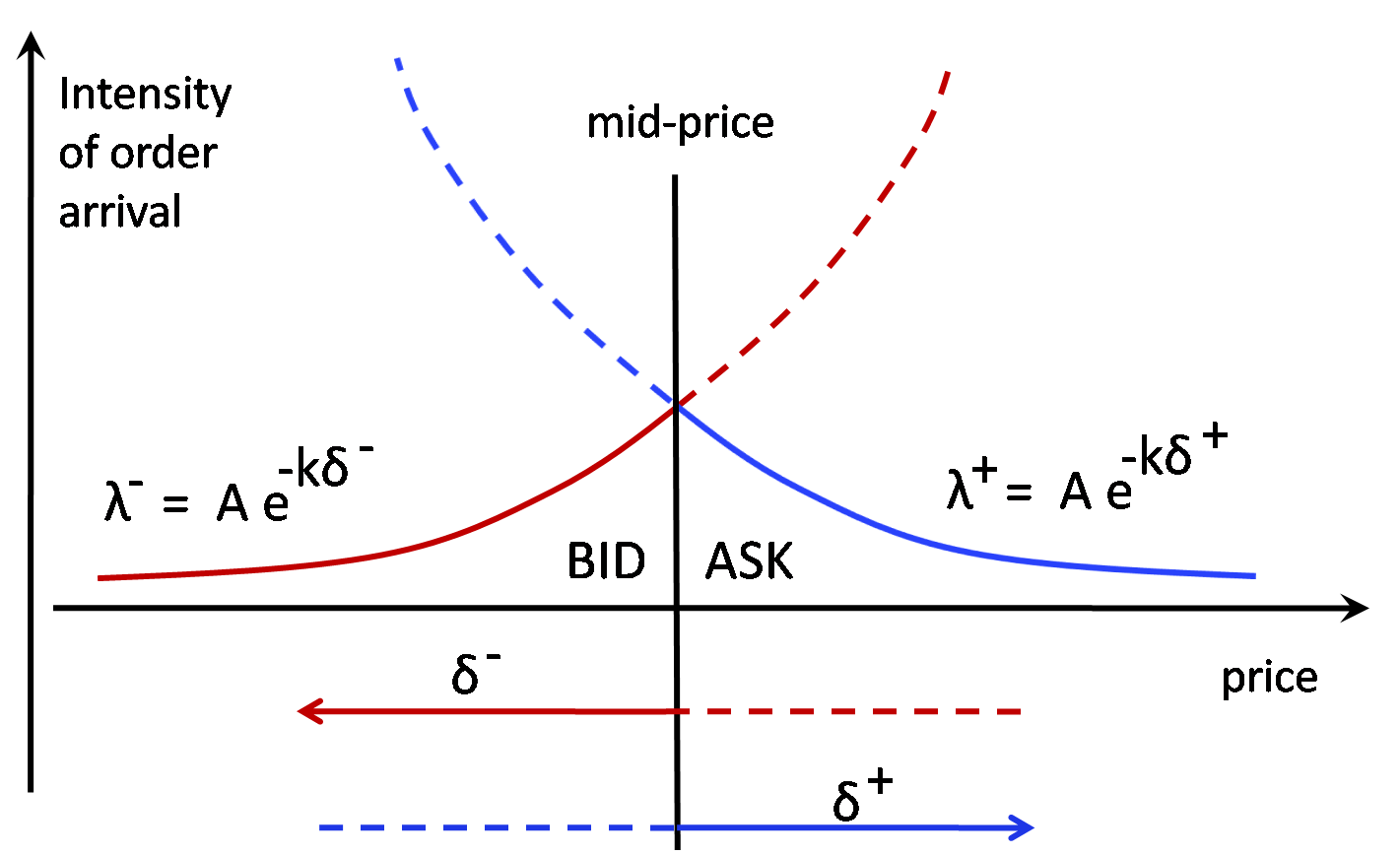}\\
Figure 2. Extrapolation of the intensities $\lambda^\pm$ of limit orders (LOs) when $\delta^\pm\le0$ (dotted lines). For our simulations, when $\delta^\pm\le0$ we will assume they are market orders.
\end{center}

In addition to the mid-price process $S(t)$ and the predictable processes $(\delta^+,\delta^-)$, we will consider two other processes. On the one hand, the inventory $Q(t)\in\Z$, which varies with the execution of the limit orders placed by the market-maker; on the other hand, the cash process $X(t)\in\R$, which also varies as the market-maker buys or sells the asset.\\

We will assume that the dynamics of $Q(t)$ and $X(t)$ are governed by 
\begin{eqnarray*}
dQ(t) &=& dN^-(t) - dN^+(t)\,,\\
dX(t) &=& \left[ S(t) + \delta^+ \right] dN^+(t) -  \left[ S(t) - \delta^- \right] dN^-(t)\,,
\end{eqnarray*}
where $N^\pm(t)$ are two independent Poisson processes of intensity $\lambda^\pm(\delta^\pm)$ and $\delta^\pm\mapsto\lambda^\pm(\delta^\pm)$ is decreasing. In that framework, the PNL or wealth of the market-maker is
\[
\mathrm{PNL}(t) = X(t) + Q(t)S(t) \in\R\,.
\]

Finally, we will also assume that the market-maker has a utility function $\phi(s,q,x)$ and an associated value function
\[
u(t,s,q,x):=\max_{(\delta^+,\delta^-)}\E_{t,s,q,x}\Big[\phi\Big(S(T),Q(T),X(T)\Big)\Big]
\]
where $t\in[0,T]$ is the current time, $s=S(t)$ is the current mid-price of the asset, $x=X(t)$ is the current cash and $q=Q(t)$ is the current inventory level.\\

It is worth to mention that all these assumptions are also present (explicitly or implicitly) in both Avellaneda and Stoikov \cite{ave} and Lehalle \emph{et al} \cite{lehalle}.

\subsection*{The Hamilton-Jacobi-Bellman equation}

Let us give a heuristic interpretation of the Hamilton-Jacobi-Bellman equation in terms of the two infinitesimal operators, one on the continuous variable $s$ and the other on the jump variables $(q,x)$, as well as of the stochastic controls $(\delta^+,\delta^-)$.\\

First, suppose $q$ and $x$ fixed. Since the continuous variable $s$ follows \eqref{eq-pricedyn}, from Feynmann-Kac representation formula we have that if $(t,s)\mapsto u_C(t,s,q,x)$ satisfies 
\begin{equation}\label{eq-conti}
\left(\partial_t+\boL\right)u_C = 0\,,\qquad \boL := b(t,s)\partial_s + \frac{1}{2}\sigma^2(t,s)\partial_{ss}\,,\qquad u_C(T,s,q,x) = \phi(s,q,x)
\end{equation}
then (see e.g. Pham \cite{pham})
\[
u_C(t,s,q,x) = \E_{t,s}\Big[\phi\Big(S(T),q,x)\Big)\Big]\,.
\]

Second, suppose now that $s$ is fixed. We  model the arrival of orders to the limit-order Book as two independent Poisson Process, $N^+$ for the ask quotes  and $N^-$ for the bid quotes , with intensity $\lambda^\pm$ (respectively).
\begin{itemize}
\item For the ask quote, we assume that $\lambda^+$ depends on the distance to the mid-price $s$, i.e. $\lambda^+ = \lambda^+(\delta^+)$. Moreover, since buying market orders favour ask quotes with the smallest spread $\delta^+$, it is natural to assume that $\delta^+\mapsto \lambda^+(\delta^+)$ is increasing, i.e. the probability of execution for the market-maker decreases as she moves her ask quote further away from the mid price.
\item Analogously, for the bid quote we have $\lambda^- = \lambda^-(\delta^-)$ and  $\delta^-\mapsto \lambda^-(\delta^-)$ increasing.
\end{itemize}

The jump variables $(q,x)$ are related via the arrival of market orders that hit the quotes of the market-maker:
\begin{itemize}
\item Suppose a buying market order of one share of $S$ hits the ask quote of the market-maker. It follows then that $x\mapsto x+(s+\delta^+)$ and $q\mapsto q-1$.
\item Analogously, if a selling market order of one share hits her bid quote then $x\mapsto x + (s-\delta^-)$ and $q\mapsto q+1$.
\end{itemize}
Under this framework, we have that the value function of the jump $(q,x)\mapsto u_D(t,s,q,x)$ satisfies
\begin{eqnarray}\label{eq-poi}
\partial_t u_D + \lambda^+(\delta^+)\left[u_D(t,s,q-1,x+(s+\delta^+))-u_D(t,s,q,x)\right] & & \\
+\lambda^-(\delta^-)\left[u_D(t,s,q+1,x-(s-\delta^-))-u_D(t,s,q,x)\right] &=& 0\,.\nonumber\\
u(T,s,q,x) &=& \phi(s,q,x)\,.\nonumber
\end{eqnarray}

Third, let us now put together the continuous and jump dynamics \eqref{eq-conti}-\eqref{eq-poi}. Suppose that the market-maker's spread quotes $(\delta^+,\delta^-)$ are known (i.e. deterministic), the value function $u(t,s,q,x)$ satisfies has the following infinitesimal generator (see e.g. Ho and Stoll \cite{ho}):
\begin{eqnarray}\label{eq-infgen}
\left(\partial_t+\boL\right)u + \lambda^+(\delta^+)\left[u(t,s,q-1,x+(s+\delta^+))-u(t,s,q,x)\right] & & \\
+\lambda^-(\delta^-)\left[u(t,s,q+1,x-(s-\delta^-))-u(t,s,q,x)\right] &=& 0\,.\nonumber\\
u(T,s,q,x) &=& \phi(s,q,x)\,.\nonumber
\end{eqnarray}

Fourth, notice that \eqref{eq-infgen} is valid only when the spread quotes $(\delta^+,\delta^-)$ are known, but in the current case they are part of the set of unknowns of the problem. In consequence, from the stochastic control theory it follows that the value function $u(t,s,q,x)$ with unknown controls $(\delta^+,\delta^-)$ is solution of the Hamilton-Jacobi-Bellman equation 
\begin{eqnarray}\label{eq-hjb1}
\left(\partial_t+\boL\right) u + \max_{\delta^+}\lambda^+(\delta^+)\left[u(t,s,q-1,x+(s+\delta^+))-u(t,s,q,x)\right] & & \\
+\max_{\delta^-}\lambda^-(\delta^-)\left[u(t,s,q+1,x-(s-\delta^-))-u(t,s,q,x)\right] &=& 0\,,\nonumber\\
u(T,s,q,x) &=& \phi(s,q,x)\,.\nonumber
\end{eqnarray}
In general we should use $sup$ instead of $max$ in \eqref{eq-hjb1}, but if we assume that $b(t,s)$ and $\sigma(t,s)$ are Lipschitz and that the jump dynamic \eqref{eq-poi} is bounded then the supremum is attained and the solution $u(t,s,q,x)$ is unique, as in the particular cases we will consider here.\\

Fifth, Avellaneda and Stoikov \cite{ave} showed that, given the empirical evidence provided by the current research in Econophysics e.g. Potters and Bouchaud \cite{jpb}, we can assume that $\lambda^\pm(\delta)=Ae^{-k\delta}$. Under this framework, the Hamilton-Jacobi-Bellman \eqref{eq-hjb1} becomes

\begin{eqnarray}\label{eq-hjb2}
\left(\partial_t+\boL\right)u + \max_{\delta^+}Ae^{-k\delta^+}\left[u(t,s,q-1,x+(s+\delta^+))-u(t,s,q,x)\right] & & \\
+\max_{\delta^-}Ae^{-k\delta^-}\left[u(t,s,q+1,x-(s-\delta^-))-u(t,s,q,x)\right] &=& 0\,,\nonumber\\
u(T,s,q,x) &=& \phi(s,q,x)\,,\nonumber
\end{eqnarray}

\subsection*{Solving the Hamilton-Jacobi-Bellman equation}

Equation \eqref{eq-hjb2} is the one we will consider in the rest of the present work. The steps to solve it are as follows:
\begin{enumerate}
\item Based on the utility function $\phi(s,q,x)$ we make an \emph{ansatz}, i.e. we guess the general form of the solution of the on Hamilton-Jacobi-Bellman equation. For example, if
\[
\phi(s,q,x) = x + \varphi(s,q)
\]
we will use
\begin{equation}\label{eq-ans}
u(t,s,q,x) = x + v(t,s,q)\,.
\end{equation}

\item We substitute the ansatz on the HJB in order to find an easier HJB equation for $v$. We use this new HJB equation to find the optimal controls $(\delta^+_\ast,\delta^-_\ast)$ that maximize the jump. With the ansatz \eqref{eq-ans} the optimal controls are
\[
\delta^+_\ast = \frac{1}{k}-s+v(t,s,q)-v(t,s,q-1)\,,\qquad \delta^-_\ast = \frac{1}{k}+s-v(t,s,q+1)+v(t,s,q)\,.
\]

\item We substitute the optimal controls on the HJB equation: the resulting equation is called the \emph{verification equation}. In our case it is
\begin{eqnarray*}
\left(\partial_t+\boL\right)v + \frac{A}{k}\left(e^{-k\delta^+_\ast}+e^{-k\delta^-_\ast}\right) &=& 0\,,\\
v(T,s,q) &=& \varphi(s,q)\,,
\end{eqnarray*}
which is highly nonlinear because $\delta^\pm_\ast = \delta^\pm_\ast (v)$.

\item We solve the verification equation via the Feynmann-Kac representation formula in order to find an explicit expression of the optimal controls. Indeed, for an equation of the form
\begin{eqnarray*}
\left(\partial_t+\boL\right)w + f(\delta^+_\ast,\delta^-_\ast,t,s,q,x) &=& 0\,,\\
w\vert_{t=T} &=& g(\delta^+_\ast,\delta^-_\ast,s,q,x)\,,
\end{eqnarray*}
where $(\delta^+_\ast,\delta^-_\ast)$ do not depend on $w$, the (unique) solution is (see e.g. Pham \cite{pham})
\[
w(t,s,q,x) = \E_{t,s,q,x}\left[\int_t^T f(\delta^+_\ast,\delta^-_\ast,\xi,S(\xi),q,x)d\xi + g(\delta^+_\ast,\delta^-_\ast,S(T),q,x)\right]\,,
\]
where $\E_{t,s,q,x}$ is the conditional expectation given $S(t)=s$, $Q(t)=q$ and $X(t)=x$
\item Alternatively, we could express the optimal quotes in terms of the market-maker's bid-ask spread 
\[
\psi_\ast := \delta^+_\ast+\delta^-_\ast
\]
and the mid-point of the spread (called the \emph{indifference price})
\[
r_\ast := \frac{1}{2}\left(p^+_\ast+p^-_\ast\right) = s + \frac{1}{2}\left(\delta^+_\ast-\delta^-_\ast\right)\,.
\]
Notice that if $\delta^+=\delta^-$ then $r_\ast(t)=s=S(t)$. Therefore, $r_\ast-s$ measures  the level of asymmetry of the quotes with respect to the mid-price $s$.
\end{enumerate}

%%%%%%%%%%
% LINEAR
%%%%%%%%%%

\section{Linear utility function}

Let us suppose that the utility function is linear, i.e.
\[
\phi(s,q,x) = x+qs\,.
\]
Then the corresponding value function is
\begin{equation}\label{eq-ulin}
u(t,s,q,x) = \max_{(\delta^+,\delta^-)}\E_{t,s,q,x}[X(T)+Q(T)S(T)]\,,
\end{equation}
where $X(t)$ is the cash process, $Q(t)$ the inventory process, $S(t)$ the price process and $\E_{t,s,q,x}$ is the conditional expectation given $S(t)=s$, $Q(t)=q$ and $X(t)=x$. Alternatively, this corresponds to choosing the final condition as $\phi(s,q,x) = x+qs$.\\

$X(T)+Q(T)S(T)$ is the final value of the market-maker's portfolio, and corresponds to the final PNL of the market-maker. Indeed, $X(T)$ the cash she holds whilst $Q(T)S(T)$ the cash value of her inventory: she holds $Q(T)$ assets and clears them at (unitary) price $S(T)$ on the close auction.

\subsection*{Ansatz}

From the final condition (i.e. the utility function) $\phi(s,q,x) = x+qs$ we will search a solution of the form
\begin{equation}\label{eq-anslin}
u(t,s,q,x) = x + \theta_0(t,s) + q\theta_1(t,s)\,.
\end{equation}
Plugging \eqref{eq-anslin} into \eqref{eq-hjb2} yields
\begin{eqnarray*}
\left(\partial_t+\boL\right)(\theta_0+q\theta_1) + \max_{\delta^+}Ae^{-k\delta^+}\left[s+\delta^+-\theta_1\right] + \max_{\delta^-}Ae^{-k\delta^-}\left[-s+\delta^-+\theta_1\right]&=& 0\,,\\
\theta_0(T,s) &=& 0\,,\nonumber\\
\theta_1(T,s) &=& s\,.\nonumber
\end{eqnarray*}

\subsection*{Computing the optimal controls}

Define
\[
f^+(\delta^+):=Ae^{-k\delta^+}\left[s+\delta^+-\theta_1\right]\,.
\]
Using Calculus we obtain that the maximum is attained at
\[
\delta^+_\ast = \frac{1}{k}-s+\theta_1\,.
\]
Analogously, if
\[
f^-(\delta^-):=Ae^{-k\delta^-}\left[-s+\delta^-+\theta_1\right]
\]
then
\[
\delta^-_\ast = \frac{1}{k}+s-\theta_1\,.
\]
In consequence, the optimal quotes $(\delta^+_\ast,\delta^-_\ast)$, spread $\psi_\ast$ and indifference price $r_\ast$ are
\[
\delta^\pm_\ast = \frac{1}{k} \pm(\theta_1-s)\,,\qquad\psi_\ast =  \delta^+_\ast+\delta^-_\ast = \frac{2}{k}\,,\qquad r_\ast = \theta_1\,.
\]

\subsection*{Solving the verification equation}

Since
\[
f^+(\delta^+_\ast) = \frac{A}{ek}e^{-k(\theta_1-s)}\,,\qquad f^+(\delta^+_\ast) = \frac{A}{ek}e^{k(\theta_1-s)}
\]
it follows that the ansatz $\theta_0+q\theta_1$ solves the verification equation
\begin{eqnarray}\label{eq-anslinver}
\left(\partial_t+\boL\right)(\theta_0+q\theta_1) + \frac{2A}{ek}\cosh\left[k(\theta_1-s)\right]&=& 0\,\\
\theta_0(T,s) &=& 0\,.\nonumber\\
\theta_1(T,s) &=& s\,.\nonumber
\end{eqnarray}

We separate \eqref{eq-anslinver} in terms of the powers of $q$, one equation for $q^0=1$ and another for $q^1=q$. With this procedure we obtain two coupled equations,
\begin{eqnarray}\label{eq-anslin0}
\left(\partial_t+\boL\right)\theta_0+ \frac{2A}{ek}\cosh\left[k(\theta_1-s)\right]&=& 0\,\\
\theta_0(T,s) &=& 0\,,\nonumber
\end{eqnarray}
and
\begin{eqnarray}\label{eq-anslin1}
\left(\partial_t+\boL\right)\theta_1&=& 0\,\\
\theta_1(T,s) &=& s\,.\nonumber
\end{eqnarray}

Applying the Feynman-Kac formula to \eqref{eq-anslin1} and recursively to \eqref{eq-anslin0} yields, respectively (see e.g. Pham \cite{pham}),
\[
\theta_1(t,s) = \E_{t,s}[S(T)]\,,\qquad \theta_0(t,s) = \frac{2A}{ek} \E_{t,s}\left\{\int_t^T\cosh\Big[k\Big(\theta_1(\xi,S(\xi))-S(\xi)\Big)\Big]d\xi\right\}\,.
\]
In consequence,
\[
\delta^\pm_\ast = \frac{1}{k}\pm\left(\E_{t,s}[S(T)]-s\right)\,,\quad \psi_\ast = \frac{2}{k}\,,\quad r_\ast = \E_{t,s}[S(T)]
\]
and
\[
u(t,s,q,x) = x + \frac{2A}{ek} \E_{t,s}\left\{\int_t^T\cosh\Big[k\Big(\theta_1(\xi,S(\xi))-S(\xi)\Big)\Big]d\xi\right\} + q\E_{t,s}[S(T)]\,.
\]
In particular, since $\cosh(\alpha)\ge1$ and $\cosh(\alpha)=1\iff \alpha=0$ we obtain that
\[
u(t,s,q,x) \ge \underline{u}(t,s,q,x):= x + \frac{2A}{ek}(T-t) + q\E_{t,s}[S(T)]
\]
and $u(t,s,q,x)=\underline{u}(t,s,q,x)\iff \theta_1(\xi,S(\xi))=S(\xi)$ for all $\xi\in[t,T]$. Since $t\in[0,T]$ is arbitrary then taking $\xi=t$ we have that $u(t,s,q,x)=\underline{u}(t,s,q,x)\iff s=\E_{t,s}[S(T)]$, i.e. if and only if $S(t)$ is a martingale.

\subsection*{Results}

Let us summarise all our findings.

\begin{theorem}\label{thm-lin}

Consider the Hamilton-Jacobi-Bellman problem
\begin{eqnarray*}
\left(\partial_t+\boL\right)u + \max_{\delta^+}Ae^{-k\delta^+}\left[u(t,s,q-1,x+(s+\delta^+))-u(t,s,q,x)\right] & & \\
+\max_{\delta^-}Ae^{-k\delta^-}\left[u(t,s,q+1,x-(s-\delta^-))-u(t,s,q,x)\right] &=& 0\,,\nonumber\\
u(T,s,q,x) &=& x+qs\,,\nonumber
\end{eqnarray*}
which corresponds to a linear utility function $\phi(s,q,x)=x+qs$, value function
\[
u(t,s,q,x) = \max_{(\delta^+,\delta^-)}\E_{t,s,q,x}[X(T)+Q(T)S(T)]
\]
and stochastic controls $(\delta^+,\delta^-)$. Then:

\Xbox{black}
{
\begin{enumerate}

\item The optimal controls $(\delta^+_\ast,\delta^-_\ast)$, spread $\psi_\ast$ and indifference price $r_\ast$ (i.e. the centre of the spread) of the market-maker are

\[
\delta^\pm_\ast = \frac{1}{k}\pm\left(\E_{t,s}[S(T)]-s\right)\,\qquad\psi_\ast = \delta^+_\ast+\delta^-_\ast = \frac{2}{k}\,,\qquad r_\ast = \E_{t,s}[S(T)]\,.
\]

\item The (unique) solution of the HJB problem is
\[
u(t,s,q,x) = x + \frac{2A}{ek} \E_{t,s}\left\{\int_t^T\cosh\Big[k\Big(\theta_1(\xi,S(\xi))-S(\xi)\Big)\Big]d\xi\right\} + q\E_{t,s}[S(T)]\,.
\]
\item The solution $u(t,s,q,x)$ is bounded from below by
\[
\underline{u}(t,s,q,x):= x + \frac{2A}{ek}(T-t) + q\E_{t,s}[S(T)]
\] 
and
\[
u(t,s,q,x)=\underline{u}(t,s,q,x)\iff s=\E_{t,s}[S(T)]\,,
\]
i.e. if and only if $S(t)$ is a martingale. In this case we have \[r_\ast(t)=S(t)\,,\qquad \delta^\pm_\ast = \frac{1}{k}\qquad\textnormal{and}\qquad u(t,s,q,x) = x+qs\,.
\]

\end{enumerate}
}

\end{theorem}

\subsection*{Remarks}

\begin{itemize}
\item The worst price dynamic for the PNL-based utility function \eqref{eq-ulin} is a martingale, in the sense that with any other price dynamic the PNL is greater. But observe that the optimal spread $\psi_\ast$ is centred around $r_\ast = \E_{t,s}[S(T)]$ and not around $s=S(t)$. In consequence, if the market-maker considers that the current mid-price $s$ has deviated from its fundamental value $r_\ast$ then she can make directional bets via her bid-ask quotes, which yields a higher PNL than the martingale assumption if the bet is correct.

\item Applying perturbation methods on the variable $q$ of the form
\[
u(t,s,q,x) = x+\theta_0(t,s)+q\theta_1(t,s)+q^2\theta_2(t,s)+\cdots\,,
\]
is a very rough approximation, to say the least. Indeed, as Lehalle \emph{et al} \cite{lehalle} pointed out, $q$ is an integer, i.e. discrete and not small, and as such a perturbation method on $q$ cannot be performed. However, once the ansatz is shown to solve the verification equation, then by uniqueness it coincides with the solution of the original problem. Therefore, the separation of the equation into two terms, one with $q^0$ and another with $q^1$, is justified \emph{a posteriori} via the maximum principle (i.e. existence and uniqueness) for the Hamilton-Jacobi-Bellman equation, and as such it does not rely at all on perturbation methods, as Avellaneda and Stoikov \cite{ave} suggested.

\item In Theorem \ref{thm-lin} we have implicitly assumed that the value function $u(t,s,q,x)$ is finite when we applied the Fenmann-Kac formula. However, this is valid if and only if
\begin{equation}\label{eq-ufin}
\E_{t,s}\left\{\int_t^T\cosh\Big[k\Big(\theta_1(\xi,S(\xi))-S(\xi)\Big)\Big]d\xi\right\}<\infty\,,\qquad \theta_1(t,s) = \E_{t,s}[S(T)]\,.
\end{equation}
If $S(t)$ is a martingale then \eqref{eq-ufin} holds trivially. For a non-martingale mid-price process $S(t)$, two sufficient conditions for \eqref{eq-ufin} to hold are $(i)$ the conditional expectation $\E_{t,s}[S(T)]$ is affine on $s$ and $(ii)$ the moment-generating function $M_Z(\lambda)=\E\left[\exp\{\lambda S(t)\}\right]$ is finite for all $\lambda\in\R$. This is the case for any Gaussian Markov process, e.g. an arithmetic Brownian motion with drift and the Ornstein-Uhlenbeck process. However, \eqref{eq-ufin} does not hold for the geometric Brownian motion with drift.

%\item We do not make any assumption on the price process $S(t)$ beyond the Feynmann-Kac representation formula. Therefore, any Markov process can be used, even processes with jumps e.g. L\'evy processes with an infinitesimal operator of the form
%\[
%\boL w = b(t,s)\partial_s w + \frac{1}{2}\sigma^2(t,s)\partial_{ss} w + \int_{-\infty}^{+\infty} \Big[w\left(s + J(s)\right)-w(s)\Big]dJ(s)\,,
%\]
%where $J(s)$ is the jump magnitude and $dJ(s)$ the jump measure. In fact, the optimal quotes are \emph{model-free} in the sense that, once we have computed or estimated $\E_{t,s}[S(T)]$ by any means, we have completely solved the optimal control problem.

\end{itemize}

%%%%%%%%%%%%%%%%%%%%%%%%%%%
% SECTION LINEAR W/ PENALTY
%%%%%%%%%%%%%%%%%%%%%%%%%%%

\section{Linear utility function with inventory penalty}

With the linear utility function there is no penalty if at the end of the trading day the market-maker carries a huge inventory. In order to force a liquidation of the inventory before the end of the day, we propose the following utility function,
\[
\phi(s,q,x) = x+qs-\eta q^2\,,\qquad \eta\ge0\,,
\]
which is the PNL with a quadratic penalty on the inventory. The associated value function is
\begin{equation}\label{eq-uquad}
u(t,s,q,x) = \max_{(\delta^+,\delta^-)}\E_{t,s,q,x}[X(T)+Q(T)S(T)-\eta Q^2(T)]\,,
\end{equation}

A quadratic penalty function for a market-maker is already known in the literature (see e.g. Stoll \cite{stoll}).

\subsection*{Ansatz}

Given the form of the utility function, we will search a solution of the form
\begin{equation}\label{eq-ansquad}
u(t,s,q,x) = x + \theta_0(t,s) + q\theta_1(t,s) - \eta q^2\theta_2(t,s)\,.
\end{equation}
Plugging \eqref{eq-ansquad} into \eqref{eq-hjb2} yields
\begin{eqnarray*}
\left(\partial_t+\boL\right)(\theta_0+q\theta_1- \eta q^2\theta_2) + \max_{\delta^+}Ae^{-k\delta^+}\left[s+\delta^+-\theta_1-\eta(1-2q)\theta_2\right]  & & \\ +\max_{\delta^-}Ae^{-k\delta^-}\left[-s+\delta^-+\theta_1-\eta(1+2q)\theta_2\right]&=& 0\,,\\
\theta_0(T,s) &=& 0\,,\nonumber\\
\theta_1(T,s) &=& s\,.\nonumber\\
\theta_2(T,s) &=& 1\,.\nonumber
\end{eqnarray*}

\subsection*{Computing the optimal controls}

As in the previous section, if
\[
f^+(\delta^+):=Ae^{-k\delta^+}\left[s+\delta^+-\theta_1-\eta(1-2q)\theta_2\right]
\]
then
\[
\delta^+_\ast  = \frac{1}{k}-s+\theta_1+\eta(1-2q)\theta_2\,.
\]
On the other hand, if
\[
f^-(\delta^-):=Ae^{-k\delta^-}\left[-s+\delta^-+\theta_1-\eta(1+2q)\theta_2\right]
\]
then
\[
\delta^-_\ast  = \frac{1}{k}+s-\theta_1+\eta(1+2q)\theta_2\,.
\]
In consequence, the optimal quotes $(\delta^+_\ast,\delta^-_\ast)$, spread $\psi_\ast$ and indifference price $r_\ast$ are
\[
\delta^\pm_\ast = \frac{1}{k}+\eta\theta_2 \pm\left( \theta_1-s-2q\eta\theta_2\right)\,,\qquad
\psi_\ast = \delta^+_\ast+\delta^-_\ast = \frac{2}{k}+2\eta\theta_2\,,\qquad r_\ast = \theta_1-2\eta q\theta_2\,.
\]

\subsection*{Solving the equation with linear jumps}

%Since
%\[
%f^+(\delta^+_\ast) = \frac{A}{k+\gamma}\exp\left\{-k\left( \frac{1}{\gamma}\log\left(1+\frac{\gamma}{k}\right)-\theta_2\right)\right\}\exp\left\{-k\left(\theta_1-s+2q\theta_2\right)\right\}
%\]
%and
%\[
%f^-(\delta^-_\ast) = \frac{A}{k+\gamma}\exp\left\{-k\left( \frac{1}{\gamma}\log\left(1+\frac{\gamma}{k}\right)-\theta_2\right)\right\}\exp\left\{+k\left(\theta_1-s+2q\theta_2\right)\right\}
%\]

We fix $q\in\Z$ and define the jump functional
\[
J_q(\delta^+_\ast,\delta^-_\ast) := \frac{A}{ek}\left(e^{1-k\delta^+_\ast}+e^{1-k\delta^-_\ast}\right)\,,
\]
whose first-order Taylor expansion (i.e. its Fr\'echet derivative) is
\begin{eqnarray*}
J_q(\delta^+_\ast,\delta^-_\ast) &=& \frac{A}{ek}\left(4-k(\delta^+_\ast+\delta^-_\ast)\right) + O\left(\vert 1-k\delta^+_\ast \vert^2+\vert 1-k\delta^-_\ast\vert^2\right)\\
 &=& \frac{A}{ek}\left(2-k\eta\theta_2 \right) + O\left(\vert 1-k\delta^+_\ast \vert^2+\vert 1-k\delta^-_\ast\vert^2\right)\,.
\end{eqnarray*}
Since at first order the jumps are independent of $q$, it follows that
\[
\theta(t,s,q) = \theta_0(t) + q\theta_1(t,s)-\eta q^2\theta_2(t)
\]
solves
\begin{eqnarray}\label{eq-linqver}
\left(\partial_t+\boL\right)\theta +\frac{A}{ek}\left(2-k\eta\theta_2 \right) &=& 0\,\nonumber\\
\theta(T,s,q) &=& qs-\eta q^2\,.\nonumber
\end{eqnarray}

We separate \eqref{eq-linqver} in terms of the powers of $q$, one equation for $q^0=1$ and another for $q^1=q$. With this procedure we obtain three coupled equations,
\begin{eqnarray}\label{eq-linq0}
\partial_t\theta_0 +\frac{A}{ek}\left(2-k\eta\theta_2 \right) &=& 0\,\\
\theta_0(T) &=& 0\,,\nonumber
\end{eqnarray}
\begin{eqnarray}\label{eq-linq1}
\left(\partial_t+\boL\right)\theta_1&=& 0\,\\
\theta_1(T,s) &=& s\,,\nonumber
\end{eqnarray}
and
\begin{eqnarray}\label{eq-linq2}
\partial_t\theta_2&=& 0\,\\
\theta_2(T) &=& 1\,.\nonumber
\end{eqnarray}

In consequence,
\[
\theta_2=1\,,\qquad\theta_1(t,s) = \E_{t,s}[S(T)]\,,\qquad \theta_0(t) = \frac{A}{ek}\left(2-k\eta\right)(T-t)\,.
\]

\subsection*{Finding a sub-solution}

Since 
\[
\frac{A}{ek}\left(e^{1-k\delta^+_\ast}+e^{1-k\delta^-_\ast}\right)\ge\frac{A}{ek}\left(4-k(\delta^+_\ast+\delta^-_\ast)\right)
\]
then the solution of the equation with linear jump (i.e. first-order Taylor) is a sub-solution of the original problem with exponential jump. In other words, if we define
\[
\underline{u}(t,s,q,x) := x+\theta_0(t)+q\theta_1(s,t)
-\eta q^2\theta_2(t)\,,
\]
where $\theta_0(t)$, $\theta_1(s,t)$ and $\theta_2(t)$ are defined as above then
\[
\underline{u}(t,s,q,x)\le u(t,s,q,x)\,,
\]
i.e. it is a sub-solution of the HJB equation.

\subsection*{Results}

Let us summarise all our findings.

\begin{theorem}\label{thm-quad}

Consider the Hamilton-Jacobi-Bellman problem
\begin{eqnarray*}
\partial_t u + \boL u + \max_{\delta^+}Ae^{-k\delta^+}\left[u(t,s,q-1,x+(s+\delta^+))-u(t,s,q,x)\right] & & \\
+\max_{\delta^-}Ae^{-k\delta^-}\left[u(t,s,q+1,x-(s-\delta^-))-u(t,s,q,x)\right] &=& 0\,,\nonumber\\
u(T,s,q,x) &=& x+qs-\eta q^2\,,\nonumber
\end{eqnarray*}
which corresponds to a linear utility function with quadratic inventory penalty $\phi(s,q,x)=x+qs-\eta q^2$, value function
\[
u(t,s,q,x) = \max_{(\delta^+,\delta^-)}\E_{t,s,q,x}\left[X(T)+Q(T)S(T)-\eta Q^2(T)\right]
\]
and stochastic controls $(\delta^+,\delta^-)$. Then:

\Xbox{black} {
\begin{enumerate}

\item Let $u(t,s,q,x)$ be the (unique) solution of the HJB equation. Then
\[
\underline{u}(t,s,q,x) := x + \frac{A}{ek}\left(2-k\eta\right)(T-t) + q\E_{t,s}[S(T)] -\eta q^2
\]
is a sub-solution of the HJB equation and $u(t,s,q,x) \ge \underline{u}(t,s,q,x)$.

\item With the linear aproximation of the jumps, or equivalently using the HJB of the sub-solution $\underline{u}$, the optimal controls $(\delta^+,\delta^-)$, spread $\psi_\ast$ and indifference price $r_\ast$ are
\[
\delta^\pm_\ast = \frac{1}{k}+\eta\pm\left( \E_{t,s}[S(T)]-s-2q\eta\right)\,,\qquad
\psi_\ast = \delta^+_\ast+\delta^-_\ast = \frac{2}{k}+2\eta\,,\qquad r_\ast = \E_{t,s}[S(T)]-2\eta q\,.
\]

\end{enumerate}
}

\end{theorem}

\subsection*{General inventory penalties}

Suppose that the utility function is now 
\[
\phi(s,q,x) = x+qs-\eta q^2\pi(s)\,,\qquad \eta\ge0\,,
\]
where $s\mapsto\pi(s)$ is continuous and for $s\ge0$ it is non-decreasing and non-negative. For example, if $\pi\equiv 1$ we recover the previous case whilst if $\pi(s)=s^2$ we recover the classical mean-variance PNL criterion. The associated value function is
\begin{equation}\label{eq-uphi}
u(t,s,q,x) = \max_{(\delta^+,\delta^-)}\E_{t,s,q,x}[X(T)+Q(T)S(T)-\eta Q^2(T)\pi(S(T))]\,.
\end{equation}
With the ansatz
\begin{equation}\label{eq-ansphi}
u(t,s,q,x) = x + \theta_0(t,s) + q\theta_1(t,s) - \eta q^2\theta_2(t,s)
\end{equation}
it can be shown that the functions $\theta_0-\theta_2$ solve the equations
\begin{eqnarray}\label{eq-psi0}
\left(\partial_t+\boL\right)\theta_0 +\frac{A}{ek}\left(2-k\eta\theta_2 \right) &=& 0\,\\
\theta_0(T) &=& 0\,,\nonumber
\end{eqnarray}
\begin{eqnarray}\label{eq-psi1}
\left(\partial_t+\boL\right)\theta_1&=& 0\,\\
\theta_1(T,s) &=& s\,,\nonumber
\end{eqnarray}
and
\begin{eqnarray}\label{eq-psi2}
\left(\partial_t+\boL\right)\theta_2&=& 0\,\\
\theta_2(T) &=& \pi(s)\,,\nonumber
\end{eqnarray}
whose explicit solutions are
\[
\theta_2(t,s)=\E_{t,s}[\pi(S(T))]\,,\quad \theta_1(t,s)=\E_{t,s}[S(T)]\,,\quad \theta_0(t,s)=\frac{2A}{ek}(T-t)-\frac{\eta A}{e}\E_{t,s}\left[\int_t^T \theta_2(\xi,S(\xi))d\xi\right]\,.
\]
In the light of these results, we have the following extension of Theorem 2 to a general penalty function $\pi$.

\begin{theorem}\label{thm-pi}

Consider the Hamilton-Jacobi-Bellman problem
\begin{eqnarray*}
\partial_t u + \boL u + \max_{\delta^+}Ae^{-k\delta^+}\left[u(t,s,q-1,x+(s+\delta^+))-u(t,s,q,x)\right] & & \\
+\max_{\delta^-}Ae^{-k\delta^-}\left[u(t,s,q+1,x-(s-\delta^-))-u(t,s,q,x)\right] &=& 0\,,\nonumber\\
u(T,s,q,x) &=& x+qs-\eta q^2\pi(s)\,,\nonumber
\end{eqnarray*}
which corresponds to a linear utility function with quadratic inventory penalty $\phi(s,q,x)=x+qs-\eta q^2\pi(s)$, value function
\[
u(t,s,q,x) = \max_{(\delta^+,\delta^-)}\E_{t,s,q,x}\left[X(T)+Q(T)S(T)-\eta Q^2(T)\pi(S(T))\right]
\]
and stochastic controls $(\delta^+,\delta^-)$. Then:

\Xbox{black} {
\begin{enumerate}

\item If $u(t,s,q,x)$ is the (unique) solution of the HJB equation then
\[
\underline{u}(t,s,q,x) := x + \frac{2A}{ek}(T-t)-\frac{\eta A}{e}\E_{t,s}\left[\int_t^T \theta_2(\xi,S(\xi))d\xi\right] + q\E_{t,s}[S(T)] -\eta q^2\theta_2(t,s)\,,\quad 
\]
where $\theta_2(t,s)=\E_{t,s}[\pi(S(T))]$, is a sub-solution of the HJB equation and $\underline{u}(t,s,q,x)\le u(t,s,q,x)$.

\item With the linear aproximation of the jumps, or equivalently  using the HJB of the sub-solution $\underline{u}$, the optimal controls $(\delta^+,\delta^-)$, spread $\psi_\ast$ and indifference price $r_\ast$ are
\[
\delta^\pm_\ast = \frac{1}{k}+\eta\pm\left( \E_{t,s}[S(T)]-s-2q\eta\E_{t,s}[\pi(S(T))]\right)\,,\qquad
\psi_\ast = \delta^+_\ast+\delta^-_\ast = \frac{2}{k}+2\eta\,,
\]
\[
r_\ast = \E_{t,s}[S(T)]-2\eta q\E_{t,s}[\pi(S(T))]\,.
\]

\end{enumerate}
}

\end{theorem}

\subsection*{Remarks}

\begin{itemize}

\item The optimal controls depend on the ansatz we make on the utility function $u(t,s,q,x)$, i.e. on the functions $\theta_1(t,s)$ and $\theta_2(t)$. However, without the linear aproximation of the jumps our ansatz cannot give the solution: in fact, there is no solution with the chosen ansatz. That said, given that the sub-solution $\underline{u}$ is explicit, and by definition it provides a lower bound on the real solution $u$, we can consider that the quotes we have found are optimal for the sub-solution, which is a lower bound on the (penalised) PNL. Under that spirit, the optimal quotes can be interpreted as \emph{conservative} estimates of the real optimal quotes, since they minimise the potential drops on the PNL.

\item If $\eta=0$ we recover the optimal controls and the sub-solution of the linear case without inventory penalty, i.e. Theorem \ref{thm-lin}. Therefore, our linear approximation of the jumps is consistent, in the sense that it provides a perturbation of the optimal quotes in terms of the "inventory-risk" or "risk-aversion" parameter $\eta$.

\item When $\eta>0$  the spread $\psi_\ast$ widens and the indifference price $r_\ast$ shifts downwards (resp. upwards) if the inventory is positive (resp. negative), which is in line with the intuition on the inventory risk. Indeed, if the net position of the market-maker is long (resp. short) then she will improve the current ask (resp. bid) quote to lure buyers (resp. sellers), and simultaneously she will try to hide her bid (resp. ask) quote deep into the limit-order Book to deter sellers (resp. buyers). By doing so, she favours the probability of being executed in the direction that makes her to go back to zero.

\item Notice that since the linear approximation of the jump functional (i.e. its Fr\'echet derivative) does not depend on $q$, the solution $\underline{u}$ of the approximate verification equation does not rely on perturbation methods and asymptotic expansions on $q$. Of course, the solution $u$ to the verification equation with exponential jumps will indeed depend on $q$, and as such an approach similar to Lehalle \emph{et al} \cite{lehalle} is needed in order to deal with the discrete variable $q$.

\item For Theorems \ref{thm-quad} and \ref{thm-pi} we are assuming that the corresponding value function $u(t,s,q,x)$ is finite in order to apply the Feynmann-Kac formula. In the current framework, given a mid-price dynamic we choose the penalty function $\pi$ such that $u(t,s,q,x)$ is finite. In that spirit, if $S(t)$ is Gaussiann (e.g. arithmetic Brownian motion or Ornstein-Ulenbeck) or a martingale (even with jumps) then $\pi\equiv1$ suffices, whilst if $S(t)$ is a geometric Brownian motion then $\pi(s)=s^2$ is a viable candidate.

\item As we mentioned above, it is easy to find sufficient conditions to ensure boundedness of the value function $u_\eta$ for $\eta>0$. If $u_0$ is the explicit solution without inventory constraints of theorem \ref{thm-lin} then the maximum principle shows that  $\underline{u}_\eta\le u_\eta\le u_0$. Therefore, if $u_0$ is finite then $u_\eta$ is bounded. This is true for processes $S(t)$ such that $\E_{t,s}[S(T)]$ is affine in $s$, i.e.
\[
\E_{t,s}[S(T)] = \alpha(t) + \beta(t)s\,,\qquad \alpha,\beta\in C^0[0,T]\,.
\] 
In particular, $u_\eta$ is bounded if $S(t)$ is an arithmetic Brownian motion or an Ornstein-Uhlenbeck process.
 
\end{itemize}

%%%%%%%%%%%%%%%%%%%%%%
% SECTION EXPONENTIAL
%%%%%%%%%%%%%%%%%%%%%%

\section{Exponential utility function}

This case has been entirely solved by Avellaneda and Stoikov \cite{ave} and Lehalle \emph{et al} \cite{lehalle} when the mid-price is a Brownian motion. In this section we show that their approach can be easily extended to several other mid-price dynamics, e.g. Ornstein-Uhlenbeck.\\

Let us suppose that the utility function is exponential $\phi(s,q,x) =-\exp\{-\gamma(x+qs)\}$, whose coresponding value function is
\begin{equation}\label{eq-exp}
u(t,s,q,x) = \max_{(\delta^+,\delta^-)}\E_{t,s,q,x}\left[-\exp\left\{-\gamma\Big(X(T)+Q(T)S(T)\Big)\right\}\right]\,,
\end{equation}

\subsection*{Ansatz}

From the form of the utility function we will search a solution of the form
\begin{equation}\label{eq-ansexp}
u(t,s,q,x) = -\exp\{-\gamma(x+\theta(t,s,q)\}\,,\qquad \theta(t,s,q) = \theta_0(t) + q\theta_1(t,s)+q^2\theta_2(t)\,.
\end{equation}
Plugging \eqref{eq-ansexp} into \eqref{eq-hjb2} yields the Hamilton-Jacobi-Bellman for $\theta(t,s,q)$, i.e.
\begin{eqnarray}\label{eq-hjbexp}
\left(\partial_t+\boL\right)\theta -\frac{1}{2}\sigma^2\gamma\left(\partial_s\theta\right)^2 + \frac{A}{\gamma}\max_{\delta^+}e^{-k\delta^+}\left[1-\exp\{-\gamma(s+\delta^+-\theta_1+(1-2q)\theta_2)\}\right] & & \nonumber\\
+  \frac{A}{\gamma}\max_{\delta^-}Ae^{-k\delta^-}\left[1-\exp\{-\gamma(-s+\delta^-+\theta_1+(1+2q)\theta_2)\}\right]]&=& 0\,,\\
\theta(T,s) &=& x+qs\,,\nonumber
\end{eqnarray}

\subsection*{Computing the optimal controls}

For the function
\[
f^+(\delta^+):=\frac{A}{\gamma}e^{-k\delta^+}\left[1-\exp\{-\gamma(s+\delta^+-\theta_1+(1-2q)\theta_2)\}\right]
\]
its maximum is attained at
\[
\delta^+_\ast = \frac{1}{\gamma}\log\left(1+\frac{\gamma}{k}\right)-s+\theta_1-(1-2q)\theta_2\,.
\]
Analogously, if
\[
f^-(\delta^-):= \frac{A}{\gamma}\max_{\delta^-}Ae^{-k\delta^-}\left[1-\exp\{-\gamma(-s+\delta^-+\theta_1+(1+2q)\theta_2)\}\right]
\]
then
\[
\delta^-_\ast = \frac{1}{\gamma}\log\left(1+\frac{\gamma}{k}\right)+s-\theta_1-(1+2q)\theta_2\,.
\]
In consequence, the optimal quotes $(\delta^+_\ast,\delta^-_\ast)$, spread $\psi_\ast$ and indifference price $r_\ast$ are
\begin{equation}\label{eq-expcontrols}
\delta^\pm_\ast = \frac{1}{\gamma}\log\left(1+\frac{\gamma}{k}\right)-\theta_2\pm(\theta_1-s+2q\theta_2)\,,\quad   \psi_\ast = \frac{2}{\gamma}\log\left(1+\frac{\gamma}{k}\right)-2\theta_2\,,\quad r_\ast = \theta_1+2q\theta_2\,.
\end{equation}

\subsection*{Solving the equation with linear jumps}

%Since
%\[
%f^+(\delta^+_\ast) = \frac{A}{k+\gamma}\exp\left\{-k\left( \frac{1}{\gamma}\log\left(1+\frac{\gamma}{k}\right)-\theta_2\right)\right\}\exp\left\{-k\left(\theta_1-s+2q\theta_2\right)\right\}
%\]
%and
%\[
%f^-(\delta^-_\ast) = \frac{A}{k+\gamma}\exp\left\{-k\left( \frac{1}{\gamma}\log\left(1+\frac{\gamma}{k}\right)-\theta_2\right)\right\}\exp\left\{+k\left(\theta_1-s+2q\theta_2\right)\right\}
%\]

For $q\in\Z$ fixed we define the jump functional as $J_q:\R^2\to\R$ as
\begin{equation}\label{eq-jump}
J_q(\delta^+_\ast,\delta^-_\ast) := \frac{A}{k+\gamma}\left(e^{-k\delta^+_\ast}+e^{-k\delta^-_\ast}\right)\,.
\end{equation}
The first-order Taylor expansion of $J_q$ (i.e. its Fr\'echet derivative) is
\begin{eqnarray*}
J_q(\delta^+_\ast,\delta^-_\ast) &=& \frac{A}{k+\gamma}\left(2-k(\delta^+_\ast+\delta^-_\ast)\right) + O\left(\vert 1-k\delta^+_\ast \vert^2+\vert 1-k\delta^-_\ast\vert^2\right)\\
 &=& \frac{2A}{k+\gamma}\left( 1-\frac{k}{\gamma}\log\left(1+\frac{\gamma}{k}\right)+k\theta_2\right) + O\left(\vert 1-k\delta^+_\ast \vert^2+\vert 1-k\delta^-_\ast\vert^2\right)\,.
\end{eqnarray*}
Therefore, at first order we have that $J_q=J_0$ for all $q\in\Z$, i.e. the jumps are independent of $q$. In consequence,
\[
\theta(t,s,q) = \theta_0(t) + q\theta_1(t,s)+q^2\theta_2(t)
\]
solves
\begin{eqnarray}\label{eq-expver}
\left(\partial_t+\boL\right)(\theta_0 + q\theta_1+q^2\theta_2) -\frac{1}{2}q^2\sigma^2\gamma\left(\partial_s\theta_1\right)^2+\frac{2A}{k+\gamma}\left( 1-\frac{k}{\gamma}\log\left(1+\frac{\gamma}{k}\right)+k\theta_2\right)&=& 0\,\nonumber\\
\theta(T,s,q) &=& qs\,.\nonumber
\end{eqnarray}

We separate \eqref{eq-expver} in terms of the powers of $q$, one equation for $q^0=1$ and another for $q^1=q$. With this procedure we obtain three coupled equations,
\begin{eqnarray}\label{eq-exp0}
\partial_t\theta_0+\frac{2A}{k+\gamma}\left( 1-\frac{k}{\gamma}\log\left(1+\frac{\gamma}{k}\right)+k\theta_2\right)&=& 0\,\\
\theta_0(T) &=& 0\,,\nonumber
\end{eqnarray}
\begin{eqnarray}\label{eq-exp1}
\left(\partial_t+\boL\right)\theta_1&=& 0\,\\
\theta_1(T,s) &=& s\,,\nonumber
\end{eqnarray}
and
\begin{eqnarray}\label{eq-exp2}
\partial_t\theta_2-\frac{1}{2}\sigma^2\gamma\left(\partial_s\theta_1\right)^2&=& 0\,\\
\theta_2(T) &=& 0\,.\nonumber
\end{eqnarray}

Applying the Feynman-Kac formula to \eqref{eq-exp1} we find
\[
\theta_1(t,s) = \E_{t,s}[S(T)]\,.
\]
Integrating \eqref{eq-exp2} we obtain
\[
-\theta_2(t) = \frac{1}{2}\gamma\int_t^T\sigma^2(\xi,S(\xi))
\Big(\partial_s\theta_1(\xi,S(\xi))\Big)^2d\xi\,.
\]
However, the ansatz we have made implies that $\theta_2$ is independent of $s$. Therefore, in order to solve \eqref{eq-exp2} we need to assume the following conditions on the price process $S(t)$:
\[
\sigma=\sigma(t)\,,\qquad \E_{t,s}[S(T)] = \alpha(t,T) + s\beta(t,T)\,.
\]
In consequence,
\[
\theta_2(t) = -\frac{1}{2}\gamma\int_t^T\sigma^2(\xi)
\beta^2(\xi,T)d\xi\,.
\]
Finally, integrating \eqref{eq-exp1} yields
\[
\theta_0(t) = \frac{2A}{k+\gamma}\left( 1-\frac{k}{\gamma}\log\left(1+\frac{\gamma}{k}\right)\right)(T-t) - \frac{k\gamma A}{k+\gamma}\int_t^T \left\{\int_\zeta^T\sigma^2(\xi)
\beta^2(\xi,T)d\xi \right\}d\zeta\,.
\]

\subsection*{Finding a sub-solution}

Since 
\[
\frac{A}{k+\gamma}\left(e^{-k\delta^+_\ast}+e^{-k\delta^-_\ast}\right)\ge\frac{A}{k+\gamma}\left(2-k(\delta^+_\ast+\delta^-_\ast)\right)
\]
then the solution of the equation with linear jump (i.e. first-order Taylor expansion)is a sub-solution of the original problem with exponential jump. In other words, if we define
\[
\underline{u}(t,s,q,x) := -\exp\left\{-\gamma\left(x+\theta_0(t)+q\theta_1(s,t)+
q^2\theta_2(t)\right)\right\}\,,
%\overline{u}(t,s,q,x) &:=& - exp\left\{-\gamma\left(x+\overline{\theta}_0(t)+q\theta_1(s,t)+
%q^2\theta_2(t)\right)\right\}\,,
\]
where $\theta_0(t)$, $\theta_1(s,t)$ and $\theta_2(t)$ are defined as above 
%\begin{eqnarray*}
%\underline{\theta}_0(t) &:=& \frac{2A}{k+\gamma}\left( 1-\frac{k}{\gamma}\log\left(1+\frac{\gamma}{k}\right)\right)(T-t) - \frac{k\gamma A}{k+\gamma}\int_t^T \left\{\int_\zeta^T\sigma^2(\xi)
%\beta^2(\xi,T)d\zeta \right\}d\xi\,,\\
%\overline{\theta}_0(t) &:=& \frac{2A}{k+\gamma}(T-t) - \frac{k\gamma A}{k+\gamma}\int_t^T \left\{\int_\zeta^T\sigma^2(\xi)
%\beta^2(\xi,T)d\zeta \right\}d\xi\,,
%\end{eqnarray*}
then
\[
\underline{u}(t,s,q,x)\le u(t,s,q,x)\,.
\]
%Moreover,  since $\theta_0$ is the only term of the ansatz that changes with a different approximation of the jump, it follows that the quotes $(\delta^+_\ast,\delta^-_\ast)$, the spread $\psi_\ast$ and the indifference price $r_\ast$ are the same for both approximations. In consequence, the optimal controls we found are consistent.

\subsection*{Adding a quadratic inventory penalty}

We modify the exponential utility function \eqref{eq-exp} by adding a quadratic inventory penalty:
\begin{equation}\label{eq-expq}
u(t,s,q,x) = \max_{(\delta^+,\delta^-)}\E_{t,s,q,x}\left[-\exp\left\{-\gamma\Big(X(T)+Q(T)S(T)-\eta Q^2(T)\Big)\right\}\right]\,.
\end{equation}

Under this new penalty framework, the computations are exactly the same as before. The only thing that changes is the equation solved by $\theta_2(t)$, i.e. 
\begin{eqnarray}\label{eq-exp2q}
\partial_t\theta_2-\frac{1}{2}\sigma^2\gamma\left(\partial_s\theta_1\right)^2&=& 0\,\\
\theta_2(T) &=& -\eta\,,\nonumber
\end{eqnarray}
whose solution is
\[
\theta_2(t) = -\eta-\frac{1}{2}\gamma\int_t^T\sigma^2(\xi)
\beta^2(\xi,T)d\xi\,.
\]
This has an impact on $\theta_0(t)$, which has the new form
\[
\theta_0(t) = \frac{2A}{k+\gamma}\left( 1-\frac{k}{\gamma}\log\left(1+\frac{\gamma}{k}\right)-k\eta\right)(T-t) - \frac{k\gamma A}{k+\gamma}\int_t^T \left\{\int_\zeta^T\sigma^2(\xi)
\beta^2(\xi,T)d\xi \right\}d\zeta\,,
\]
as well as on the optimal quotes, spread and indifference price \eqref{eq-expcontrols}.

\subsection*{Results}

Let us summarise all our findings.

\begin{theorem}\label{thm-exp}

Consider the Hamilton-Jacobi-Bellman problem
\begin{eqnarray*}
\partial_t u + \boL u + \max_{\delta^+}Ae^{-k\delta^+}\left[u(t,s,q-1,x+(s+\delta^+))-u(t,s,q,x)\right] & & \\
+\max_{\delta^-}Ae^{-k\delta^-}\left[u(t,s,q+1,x-(s-\delta^-))-u(t,s,q,x)\right] &=& 0\,,\nonumber\\
u(T,s,q,x) &=& -\exp\{-\gamma(x+qs-\eta q^2)\}\,,\nonumber
\end{eqnarray*}
which corresponds to an exponential utility function $\phi(s,q,x)=-\exp\{-\gamma(x+qs-\eta q^2)\}$, value function
\[
u(t,s,q,x) = \max_{(\delta^+,\delta^-)}\E_{t,s,q,x}\left[-\exp\left\{-\gamma\Big(X(T)+Q(T)S(T)-\eta Q^2(T)\Big)\right\}\right]\,,
\]
and stochastic controls $(\delta^+,\delta^-)$. Assume further that the mid-price process
\[
dS(t)=b(t,S(t))dt +\sigma(t)dW(t)
\]
satisfies
\[
\E_{t,s}[S(T)] = \alpha(t,T) + s\beta(t,T)\,,\qquad \forall(t,s)\,.
\]
Then:

\Xbox{black}{
\begin{enumerate}
\item With the linear approximation of the jumps, or equivalently using the HJB of the sub-solution $\underline{u}$, the optimal controls  $(\delta^+,\delta^-)$, spread $\psi_\ast$ and indifference price $r_\ast$ are
\[
\delta^\pm_\ast = \frac{1}{\gamma}\log\left(1+\frac{\gamma}{k}\right)-\theta_2\pm(\theta_1-s+2q\theta_2)\,,\quad   \psi_\ast = \frac{2}{\gamma}\log\left(1+\frac{\gamma}{k}\right)-2\theta_2\,,\quad r_\ast = \theta_1+2q\theta_2\,,
\]
where
\begin{eqnarray*}
\theta_0(t) &=& \frac{2A}{k+\gamma}\left( 1-\frac{k}{\gamma}\log\left(1+\frac{\gamma}{k}\right)-k\eta\right)(T-t) - \frac{k\gamma A}{k+\gamma}\int_t^T \left\{\int_\zeta^T\sigma^2(\xi)
\beta^2(\xi,T)d\zeta \right\}d\xi\,,\\
\theta_1(t,s) &=& \E_{t,s}[S(T)]\,,\\
\theta_2(t) &=& -\eta-\frac{1}{2}\gamma\int_t^T\sigma^2(\xi)
\beta^2(\xi,T)d\xi\,.
\end{eqnarray*}

\item Let $u(t,s,q,x)$ be the (unique) solution of the HJB equation. Then
\[
\underline{u}(t,s,q,x) := -\exp\left\{-\gamma\Big(x+\underline{\theta}(t,s,q)\Big)\right\}\,,\qquad \underline{\theta}(t,s,q) = \theta_0(t)+q\theta_1(t,s)+q^2\theta_2(t)
\]
is a sub-solution of the HJB equation and $u(t,s,q,x)\ge \underline{u}(t,s,q,x)$.

\end{enumerate}

}

\end{theorem}

\subsection*{Remarks}

\begin{itemize}

\item The linear approximation of the jumps turns out to be independent of $q$, although for higher orders this is no longer true. This means that we are not performing perturbation methods and asymptotic expansions on the (discrete) variable $q$ when we solved the approximate verification equation. Of course, if we want to solve the real HJB problem then we need to take into account the discrete variable $q$, which leads to an infinite system of equations as in Lehalle \emph{et al} \cite{lehalle}.

\item In the case of $\eta=0$ the inventory penalty tends to zero as $t\to T$. This implies that the penalisation is not stong enough to force the market-maker to finish her day with a flat inventory, as it will be shown in the numerical simulations. Therefore, it was necesary to add an "inventory-risk" parameter $\eta>0$ in order to ensure a flat inventory at the end of the day.

\end{itemize}

\subsection*{Examples}

\begin{itemize}

\item If $S(t)$ is an \textbf{arithmetic Brownian motion with drift}, i.e.
\[
dS(t) = bdt+\sigma dW(t)
\]
then
\[
\E_{t,s}[S(T)]=s+b(T-t)\,,
\]
which implies that
\[
\theta_1 = s+b(T-t)\,,\qquad \theta_2 = -\eta-\frac{1}{2}\gamma\sigma^2(T-t)\,.
\]
In consequence, the optimal controls are

\Xbox{black}{
\begin{eqnarray*}
\delta^\pm_\ast &=& \frac{1}{\gamma}\log\left(1+\frac{\gamma}{k}\right)+\eta+\frac{1}{2}\gamma\sigma^2(T-t)\pm\Big(b(T-t)-q[2\eta+\gamma\sigma^2(T-t)]\Big)\,,\\ 
\psi_\ast &=& \frac{2}{\gamma}\log\left(1+\frac{\gamma}{k}\right)+2\eta+\gamma\sigma^2(T-t)\,,\\
r_\ast &=& s+b(T-t)-q\Big(2\eta+\gamma\sigma^2(T-t)\Big)\,.
\end{eqnarray*}
In particular, if $b=\eta=0$ we recover the results of Avellaneda and Stoikov \cite{ave}.}

\item If $S(t)$ is an \textbf{Ornstein-Uhlenbeck process}, i.e. 
\[
dS(t) = a(\mu-S(t))dt+\sigma dW(t)
\]
then
\[
\E_{t,s}[S(T)]=se^{-a(T-t)}+\mu\left(1-e^{-a(T-t)}\right)\,,
\]
which implies that
\[
\theta_1 = se^{-a(T-t)}+\mu\left(1-e^{-a(T-t)}\right)\,,\qquad \theta_2 = -\eta-\frac{\gamma\sigma^2}{4a}\left(1-e^{-2a(T-t)}\right)\,.
\]
In consequence, the optimal controls are

\Xbox{black}{
\begin{eqnarray*}
\delta^\pm_\ast &=& \frac{1}{\gamma}\log\left(1+\frac{\gamma}{k}\right)+\eta+\frac{\gamma\sigma^2}{4a}\left(1-e^{-2a(T-t)}\right)\\
 &\pm& \left((\mu-s)\left(1-e^{-a(T-t)}\right)-q\left[2\eta+\frac{\gamma\sigma^2}{2a}\left(1-e^{-2a(T-t)}\right)\right]\right)\,,\\ 
\psi_\ast &=& \frac{2}{\gamma}\log\left(1+\frac{\gamma}{k}\right)+2\eta+\frac{\gamma\sigma^2}{2a}\left(1-e^{-2a(T-t)}\right)\,,\\
r_\ast &=& se^{-a(T-t)}+\mu\left(1-e^{-a(T-t)}\right)-q\left(2\eta+\frac{\gamma\sigma^2}{2a}\left(1-e^{-2a(T-t)}\right)\right)\,.
\end{eqnarray*}
}
\end{itemize}

\subsection*{Solving the nonlinear equation: Lehalle's approach}

Assuming that $S(t)$ is a Brownian motion, Lehalle \emph{et al} \cite{lehalle} found that the solution $u(t,s,q,x)$ of the nonlinear problem can be explicitly found, but that study can be easily extended to arithmetic Brownian motions with time-dependent volatility and drift.\\

\Xbox{black}{
Suppose that the mid-price process is of the form
\[
dS(t) = b(t)dt + \sigma(t)dW(t)\,,
\]
where $b(t)$ and $\sigma(t)$ are uniformly bounded in $[0,T]$. Let us make the ansatz
\begin{equation}\label{eq-ans-lehalle}
u(t,s,q,x) = -\exp\{-\gamma(x+qs)\}v_q(t)^{-\gamma/k}\,;\quad v_q\in C^1(0,T)\,,\quad q\in\Z\,.
\end{equation}
Following  Lehalle \emph{et al} \cite{lehalle} it can be shown that the optimal quotes are
\begin{eqnarray*}
\delta^+_\ast &=&  \frac{1}{\gamma}\log\left(1+\frac{\gamma}{k}\right)+\frac{1}{k}\log\left(\frac{v_q(t)}{v_{q-1}(t)}\right)\,,\\
\delta^-_\ast &=&  \frac{1}{\gamma}\log\left(1+\frac{\gamma}{k}\right)-\frac{1}{k}\log\left(\frac{v_{q+1}(t)}{v_q(t)}\right)\,,
\end{eqnarray*}
where $(v_q(t))_{q\in\Z}$ solves the (infinite) ODE system
\begin{eqnarray}\label{eq-ode-bm}
v_q^\prime(t) &=& \left(\frac{k\gamma q^2}{2}\sigma^2(t)-\gamma q b(t)\right)v_q(t)-\left(A\left(1+\frac{k}{\gamma}\right)^{-1-k/\gamma}\right)\Big(v_{q+1}(t) + v_{q-1}(t)\Big)\,, \\
v_q(T) &=& 1\,.\nonumber
\end{eqnarray}
}

Lehalle \emph{et al} \cite{lehalle} used a constructive proof to show that \eqref{eq-ode-bm} has a unique, strictly positive solution in $C^\infty\Big([0,T));\ell^2(\Z)\Big)$. However, this result can be easily proven in a non-constructive fashion.\\

Let $\bfE$ be a Banach space and consider a system on $\bfE$ of the form
\begin{eqnarray}\label{eq-ode-lip}
v_q^\prime(t) &=& F_q(t,v(t))\,, \\
v_q(T) &=& v_{q,T}>0\,, \nonumber
\end{eqnarray}
where $F(t,v)=(F_q(t,v))_{q\in\Z}:[0,T]\times\bfE\to\bfE$ is Lipshitz in $v$ uniformly in $t\in[0,T]$. Applying the Cauchy-Picard Theorem and the maximum principle for ODEs on the Banach space $C^1\Big([0,T);\bfE\Big)$ yield existence, uniqueness and positivity of the solution $v$ of \eqref{eq-ode-bm} (see e.g. Br\' ezis \cite{bre}). Unfortunately, since the linear function $F_q(t,v)$ corresponding to \eqref{eq-ode-bm} is proportional to $q^2$, we cannot apply the Cauchy-Picard Theorem directly. However, if we define
\begin{equation}\label{eq-banach}
\bfE = \ell^\infty(\Z) \,,\qquad \bfF := \left\{v\in\bfE: \sup_{q\in\Z} q^2v_q<+\infty\right\}\,,
\end{equation}
then for $t$ fixed we have that $F(t,\cdot):\bfF\to\bfE$ is linear and bounded, and the bound is uniform in $t$. Therefore, $F(t,v)$ is Lipschitz in $v$, uniformly in $t$. In consequence, we can now apply the Cauchy-Picard Theorem to ensure that there exists a unique positive solution of \eqref{eq-ode-bm}.

\subsection*{Remarks on Lehalle's approach}

\begin{itemize}

\item If instead of the Banach spaces in \eqref{eq-banach} we use the Hilbert spaces
\begin{equation}\label{eq-hilbert}
\bfE = \ell^2(\Z) \,,\qquad \bfF := \left\{v\in\bfE: \sum_{q\in\Z} q^2v_q<+\infty\right\}\,,
\end{equation}
we recover the framework of Lehalle \emph{et al} \cite{lehalle}.

\item Our proof is non-constructive, which implies that we cannot provide explicit asymptotic estimates of the solution $v=(v_q)_{q\in\Z}$. Lehalle \emph{et al} \cite{lehalle}, on the contrary, constructed the operator explicitly, and thus they were able to show the asymptotic behaviour of $v$ based on the spectrum of the linear operator.

\item From the ansatz \eqref{eq-ans-lehalle} we see that the coefficients in the ODE system \eqref{eq-ode-bm} cannot depend on $s$. This rules out mid-price processes whose drift and volatility depend on $s$, e.g. Ornstein-Uhlenbeck and geometric Brownian Motion. However, arithmetic Brownian motions with time-dependent drift and volatility can be used.

\end{itemize}

%%%%%%%%%%%%%%%%%%%%%%%
% NUMERICAL SIMULATIONS
%%%%%%%%%%%%%%%%%%%%%%%

\section{Numerical Simulations and sample paths}

We performed several simulations of the optimal market-making strategy, i.e. the spread $\psi_\ast$ and the indifference price $r_\ast$, under an Ornstein-Uhlenbeck mid-price process, i.e. a mean-reverting price dynamic of the form
\[
dS(t) = a(\mu-S(t))dt+\sigma dW(t)\,.
\]
We considered four strategies: linear/exponential utility and mean-reverting /martingale market-making assumption. This allows us to assess the effect of the directional bet $\mu$ on the PNL of the market-making strategy. The parameters we used are $k=100$, $A=1500$, $T=1$ (which corresponds to one trading day), $n=1000$ (which corresponds to 1000 bid/ask limit orders sent per day, approx. once every 30 seconds), $S(0)=1$, $\sigma = 0.05$ (daily volatility of 5\%), $a=1$ and $\eta = 0.0001$ (which is a very small inventory-risk aversion but enough to force the inventory to end the day flat in average). Under these parameters we have a constant linear spread of $\psi_\ast=0.0202$ whilst the exponential spread is time-dependent.\\

We chose three different values of $\mu$: $0.98$, $1.00$ and $1.02$. This corresponds, respectively, to a bet that the price will go down by 2\%, will oscillate around its open price or go up by 2\%. We used the linear utility function with inventory penalty and two strategies, a mean-reverting strategy (Ornstein-Uhlenbeck) with the correct directional bet, and the martingale strategy (arithmetic Brownian motion) with no directional bet. The martingale strategy performs a pure market-making strategy under inventory constraints. On the other hand, the mean-reverting strategy performs the same market-market strategy than the martingale but it also places directional bets, which can be seen not only in the agressiveness of the ask and bid quotes but also in the fact that the algorithm can place market orders.

\begin{center}
\includegraphics[width=6.5in, height=4in]{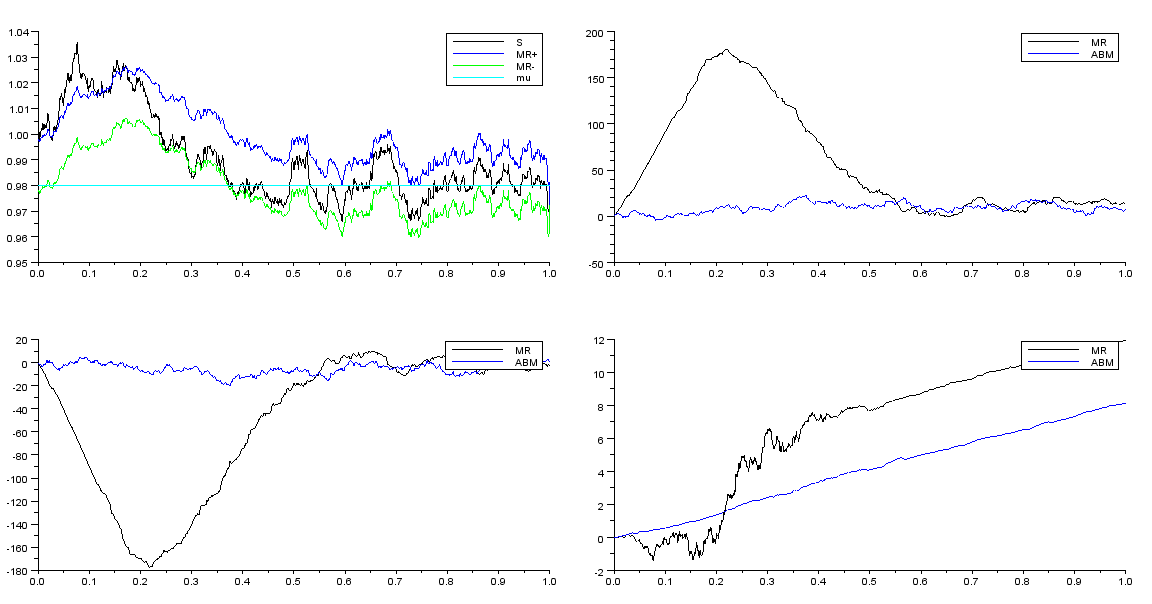}\\
Figure 3. Simulation of the market-making strategy under a mean-reverting mid-price dynamic with (asymptotic) mean $\mu=0.98$. \textbf{Upper-Left}: mid-price (black), optimal ask quote for the market-maker (dark blue), optimal bid quote (light green), $\mu$ (light blue). \textbf{Lower-Left}: Inventory for the mean-reverting process (black) vs the inventory for the martingale (blue). \textbf{Upper-Right}: cash. \textbf{Lower-Right}: PNL of the mean reverting process (black) compared with the benchmark, i.e. the PNL of the martingale strategy (blue).
\end{center}

In \textbf{Figure 3} we have plotted a realisation of the market-making strategy for $\mu=0.98$, i.e. assuming that the price will go down by 2\% at the end of the day. If the market mid-price --black line-- is above (resp. below) the optimal ask quote --dark blue-- (resp. the optimal bid quote --light green--) then the market-maker sells (resp. buys) at market price, which we assume to coincide with the mid-price.

\begin{itemize}

\item The mid-price starts at $s=1.00$, it goes up to $s=1.03$ at $t=0.1$ and stays above $s=1.01$ up to $t=0.2$. Since the bet is that the price will converge down to $\mu=0.98$, the market-maker sells the asset at market price. At $t=0.2$ the market-maker has an inventory of $q=-170$ and a PNL below the martingale benchmark: as she sold her assets at the market price (mid-price), she paid the spread to mount her directional bet.

\item During the time interval $(0.20,0.50)$ the market-maker buys back her position via limit-orders. In order to favour the arrival of selling orders and deter buying orders she plays very aggressive bid quotes and very conservative ask quotes, hence the mid-price is closer to her bid quote than her ask quotes. The strategy paid well because at $t=0.5$ the mid-price converged to $\mu=0.98$, her inventory went back to zero and her PNL outperformed the martingale benchmark.

\item On $(0.50,1.00)$ the market-maker does not make any directional bet, she only plays the bid-ask spread because the mid-price oscillates around $\mu=0.98$. As it can be seen, her bid and ask quotes are rather symmetric with respect to the mid-price, i.e. during all this non-directional period the strategy makes the same PNL than the martingale benchmark because both lines are almost parallel.

\end{itemize}

\begin{center}
\includegraphics[width=6.5in, height=4in]{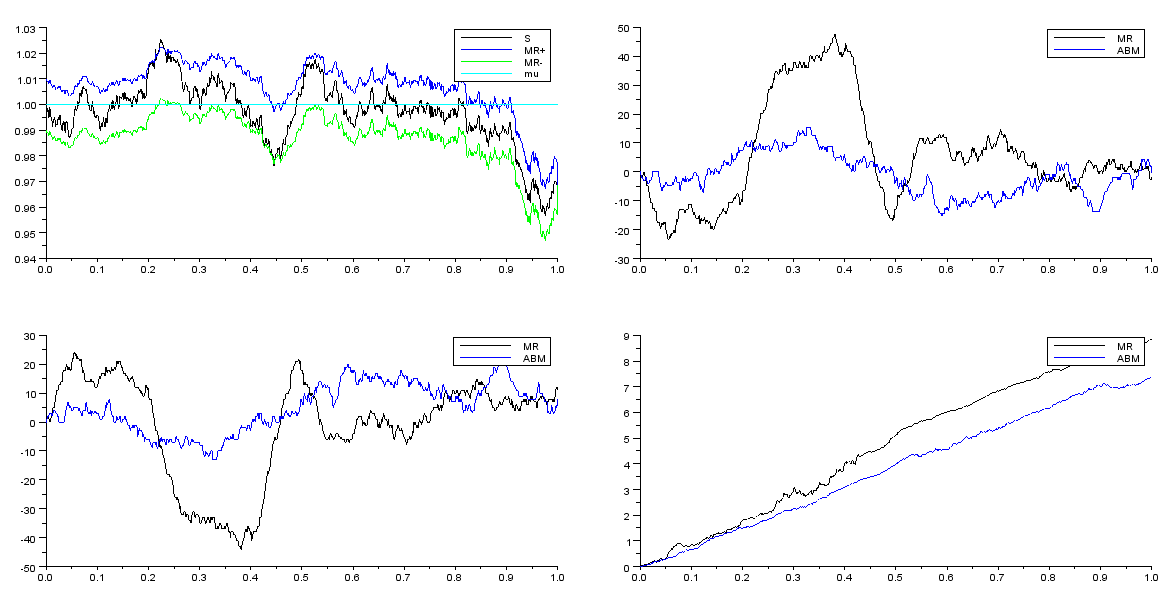}\\
Figure 4. Simulation for $\mu=1.00$.
\end{center}

In \textbf{Figure 4} the market-maker assumes that the price will oscillate around $\mu=1.00$. Therefore, her quotes are symmetric near this threshold and are tilted when the prices wander far from it (i.e. she makes mean-reverting bets). Therefore, her inventory oscillates from positive on $(0.00,0.2)$ to negative on $(0.25,0.45)$, then back to positive and negative again. On $(0.85,1.00)$ there is a huge drop in the mid-price, which in absence of inventory risk would imply a consequent positive inventory due to the mean-reverting dynamic. However, the inventory-risk-aversion forces her to avoid a directional bet at the end of the day, and as such her strategy is similar to the martingale case. It is worth to mention that the mid-price finished the day below the target of $1.00$. In consequence, the market-making strategy is insensitive to peaks at the end of the day.

\begin{center}
\includegraphics[width=6.5in, height=4in]{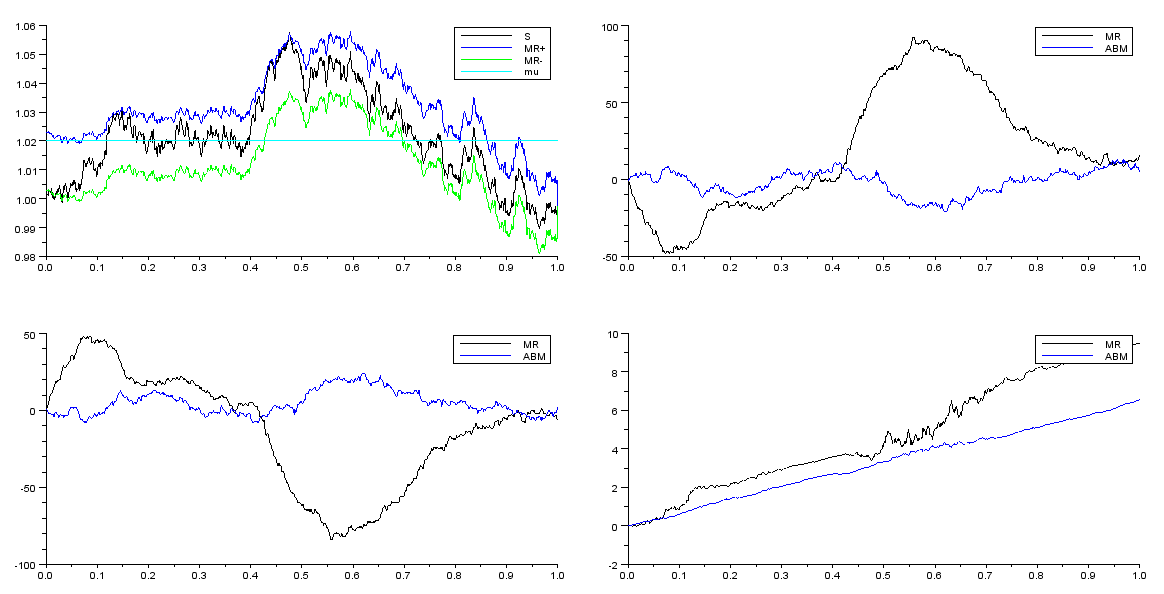}\\
Figure 5. Simulation for $\mu=1.02$.
\end{center}

In \textbf{Figure 5} the market-maker assumes that the price will converge to $\mu=1.02$. On $(0.00,0.10)$ she mounts an inventory of $q=50$ because she bets the mid-price will hit $\mu=1.02$. She eliminates her inventory during $(0.20,0.40)$ using limit-orders with generous ask prices and not-very competitive bid prices, which translates into a greater flow of buying orders than selling orders. At $t=0.20$ we can see that her directional bet has beaten the martingale benchmark. On $(0.40,0.70)$ the mid-price rises from $\mu=1.02$ and comes back, and since the market-maker is betting for a mean-reverting dynamic she builds up an inventory of $q=-90$ at $t=0.55$. On $(0.85,1.00)$ the mid-price falls, but instead of making a U-turn in her inventory and turn it positive, as a mean-reverting dynamic suggests, she rather eliminates slowly her negative inventory by tilting her quotes towards the buying side. By doing so the market-maker avoids directional bets and minimises her market impact at the end of the trading day.

\section{Statistics of the PNL distributions}

\subsection*{Comparing linear and exponential utility strategies}

We performed 100,000 simulations with $\eta\in\{0,0.0001,0.001\}$, $\gamma=1$ and $\mu\in\{0.98,1.00,1.02\}$. The rest of the parameters have the same values as for the previous simulations.

\begin{center}
\begin{tabular}{|c|c|c|c|c|c|}
\hline
 & &\textbf{linear ABM}&\textbf{linear MR}&\textbf{exp ABM} &\textbf{exp MR}\\
	\hline
\textbf{PNL} &\textbf{mean}	&11.039	&14.290	&10.668	&11.084\\
 &\textbf{std dev}	&1.013	&13.678	&0.356	&0.520\\
 &\textbf{"Sharpe"}	&10.90	&1.04	&29.97	&21.32\\
 &\textbf{skewness}	&0.075	&-0.550	&0.026	&-1.008\\
 &\textbf{kurtosis}	&5.721	&4.692	&3.043	&6.176\\
 &\textbf{Jarque Bera}	&30947.9	&16970.7	&19.1	&58949.6\\
 &\textbf{VaR 5\%}	&9.430	&-10.909	&10.086	&10.177\\
 &\textbf{VaR 1\%}	&8.346	&-25.726	&9.842	&9.465\\
\hline
\textbf{Inv} &\textbf{mean}	&0.087	&-333.297	&0.004	&-3.311\\
 &\textbf{std dev}	&33.258	&418.200	&7.672	&15.227\\
 &\textbf{skewness}	&-0.005	&0.622	&0.003	&0.007\\
 &\textbf{kurtosis}	&2.981	&2.346	&3.001	&3.010\\
 &\textbf{Jarque Bera}	&1.8	&8232.0	&0.2	&1.3\\
 &\textbf{Q(T) 90\%}	&[-55,55]	&[-847,463]	&[-13,13]	&[-28,22]\\
\hline
\end{tabular}

\smallskip

Table 1A. Statistics of the PNL distribution. $\mu=0.98$, $\eta=0$. ABM = arithmetic Brownian Motion without drift, MR = mean-reverting. The "Sharpe" value is the normalised return per risk unit, i.e. mean / std dev. 
\end{center}

\bigskip

\begin{center}
\begin{tabular}{|c|c|c|c|c|c|}
\hline
 & &\textbf{linear ABM}&\textbf{linear MR}&\textbf{exp ABM} &\textbf{exp MR}\\
	\hline
\textbf{PNL} &\textbf{mean}	&10.982	&11.576	&10.607	&10.945\\
 &\textbf{std dev}	&0.412	&1.541	&0.347	&0.444\\
 &\textbf{"Sharpe"}	&26.66	&7.51	&30.57	&24.65\\
 &\textbf{skewness}	&0.023	&-2.164	&0.038	&-0.605\\
 &\textbf{kurtosis}	&3.016	&11.418	&3.007	&4.460\\
 &\textbf{Jarque Bera}	&10.2	&373263.9	&24.7	&14982.6\\
 &\textbf{VaR 5\%}	&10.308	&8.640	&10.041	&10.183
\\
 &\textbf{VaR 1\%}	&10.026	&5.794	&9.809	&9.691\\
\hline
\textbf{Inv} &\textbf{mean}	&-0.020	&-1.739	&0.008	&-0.785\\
 &\textbf{std dev}	&5.025	&8.428	&4.574	&5.696\\
 &\textbf{skewness}	&0.006	&0.004	&-0.001	&-0.003\\
 &\textbf{kurtosis}	&3.024	&2.995	&2.965	&3.000\\
 &\textbf{Jarque Bera}	&3.0	&0.4	&5.2	&0.2\\
 &\textbf{Q(T) 90\%}	&[-8,8]	&[-16,12]	&[-8,8]	&[-10,9]\\
\hline
\end{tabular}

\smallskip

Table 1B. Statistics of the PNL distribution. $\mu=0.98$, $\eta=0.0001$.
\end{center}

\bigskip

\begin{center}
\begin{tabular}{|c|c|c|c|c|c|}
\hline
 & &\textbf{linear ABM}&\textbf{linear MR}&\textbf{exp ABM} &\textbf{exp MR}\\
	\hline
\textbf{PNL} &\textbf{mean}	&10.435	&10.494	&10.000	&10.234\\
 &\textbf{std dev}	&0.336	&0.364	&0.324	&0.342\\
 &\textbf{"Sharpe"}	&31.06	&28.83	&27.47	&29.92\\
 &\textbf{skewness}	&0.023	&-0.158	&0.037	&-0.034\\
 &\textbf{kurtosis}	&3.013	&3.306	&3.021	&3.073\\
 &\textbf{Jarque Bera}	&9.5	&806.5	&24.3	&41.2\\
 &\textbf{VaR 5\%}	&9.886	&9.889	&9.470	&9.672\\
 &\textbf{VaR 1\%}	&9.659	&9.594	&9.252	&9.426\\
\hline
\textbf{Inv} &\textbf{mean}	&-0.001	&-0.018	&0.005	&-0.020\\
 &\textbf{std dev}	&1.667	&1.680	&1.674	&1.673\\
 &\textbf{skewness}	&-0.011	&-0.014	&-0.010	&-0.001\\
 &\textbf{kurtosis}	&3.004	&3.030	&3.022	&3.029\\
 &\textbf{Jarque Bera}	&2.2	&7.1	&3.6	&3.5\\
 &\textbf{Q(T) 90\%}	&[-3,3]	&[-3,3]	&[-3,3]	&[-3,3]\\
\hline
\end{tabular}

\smallskip

Table 1C. Statistics of the PNL distribution. $\mu=0.98$, $\eta=0.001$.
\end{center}

\bigskip

\begin{center}
\begin{tabular}{|c|c|c|c|c|c|}
\hline
 & &\textbf{linear ABM}&\textbf{linear MR}&\textbf{exp ABM} &\textbf{exp MR}\\
	\hline
\textbf{PNL} &\textbf{mean}	&11.030	&13.457	&10.670	&11.039\\
 &\textbf{std dev}	&0.992	&11.708	&0.356	&0.501\\
 &\textbf{"Sharpe"}	&11.12	&1.15	&29.97	&22.03\\
 &\textbf{skewness}	&0.014	&-0.676	&0.022	&-0.864\\
 &\textbf{kurtosis}	&5.666	&5.176	&2.996	&5.632\\
 &\textbf{Jarque Bera}	&29607.8	&27355.7	&8.4	&41315.1\\
 &\textbf{VaR 5\%}	&9.460	&-8.180	&10.086	&10.167\\
 &\textbf{VaR 1\%}	&8.368	&-22.205	&9.850	&9.524\\
\hline
\textbf{Inv} &\textbf{mean}	&-0.021	&0.055	&-0.013	&0.031\\
 &\textbf{std dev}	&33.126	&468.213	&7.682	&15.245\\
 &\textbf{skewness}	&-0.009	&0.000	&0.007	&0.007\\
 &\textbf{kurtosis}	&2.990	&1.849	&3.024	&2.988\\
 &\textbf{Jarque Bera}	&1.8	&5523.2	&3.2	&1.5\\
 &\textbf{Q(T) 90\%}	&[-55,55]	&[-732,731]	&[-13,13]	&[-25,25]\\
\hline
\end{tabular}

\smallskip

Table 2A. Statistics of the PNL distribution. $\mu=1.00$, $\eta=0$.
\end{center}

\bigskip

\begin{center}
\begin{tabular}{|c|c|c|c|c|c|}
\hline
 & &\textbf{linear ABM}&\textbf{linear MR}&\textbf{exp ABM} &\textbf{exp MR}\\
	\hline
\textbf{PNL} &\textbf{mean}	&10.982	&11.453	&10.606	&10.913\\
 &\textbf{std dev}	&0.411	&1.317	&0.345	&0.433\\
 &\textbf{"Sharpe"}	&26.72	&8.70	&30.74	&25.20\\
 &\textbf{skewness}	&0.029	&-2.070	&0.036	&-0.554\\
 &\textbf{kurtosis}	&3.031	&10.934	&2.992	&4.443\\
 &\textbf{Jarque Bera}	&17.7	&333649.4	&22.1	&13802.7\\
 &\textbf{VaR 5\%}	&10.312	&8.956	&10.043	&10.178\\
 &\textbf{VaR 1\%}	&10.037	&6.601	&9.813	&9.699\\
\hline
\textbf{Inv} &\textbf{mean}	&-0.006	&-0.046	&0.016	&-0.018\\
 &\textbf{std dev}	&5.034	&8.378	&4.570	&5.664\\
 &\textbf{skewness}	&-0.003	&0.006	&0.008	&-0.001\\
 &\textbf{kurtosis}	&2.992	&2.974	&3.008	&2.969\\
 &\textbf{Jarque Bera}	&0.5	&3.4	&1.5	&4.0\\
 &\textbf{Q(T) 90\%}	&[-8,8]	&[-14,14]	&[-7,8]	&[-9,9]\\
\hline
\end{tabular}

\smallskip

Table 2B. Statistics of the PNL distribution. $\mu=1.00$, $\eta=0.0001$.
\end{center}

\bigskip

\begin{center}
\begin{tabular}{|c|c|c|c|c|c|}
\hline
 & &\textbf{linear ABM}&\textbf{linear MR}&\textbf{exp ABM} &\textbf{exp MR}\\
	\hline
\textbf{PNL} &\textbf{mean}	&10.436	&10.483	&10.000	&10.226\\
 &\textbf{std dev}	&0.335	&0.356	&0.325	&0.338\\
 &\textbf{"Sharpe"}	&31.15	&29.45	&30.77	&30.25\\
 &\textbf{skewness}	&0.031	&-0.097	&0.031	&-0.023\\
 &\textbf{kurtosis}	&3.001	&3.162	&2.988	&3.069\\
 &\textbf{Jarque Bera}	&15.5	&266.4	&16.5	&28.6\\
 &\textbf{VaR 5\%}	&9.888	&9.889	&9.469	&9.668\\
 &\textbf{VaR 1\%}	&9.658	&9.623	&9.246	&9.431\\
\hline
\textbf{Inv} &\textbf{mean}	&-0.011	&-0.009	&0.005	&0.006\\
 &\textbf{std dev}	&1.673	&1.684	&1.672	&1.676\\
 &\textbf{skewness}	&0.006	&-0.004	&-0.001	&0.002\\
 &\textbf{kurtosis}	&3.026	&2.992	&3.025	&3.026\\
 &\textbf{Jarque Bera}	&3.6	&0.5	&2.7	&3.0\\
 &\textbf{Q(T) 90\%}	&[-3,3]	&[-3,3]	&[-3,3]	&[-3,3]\\
\hline
\end{tabular}

\smallskip

Table 2C. Statistics of the PNL distribution. $\mu=1.00$, $\eta=0.001$.
\end{center}

\bigskip

\begin{center}
\begin{tabular}{|c|c|c|c|c|c|}
\hline
 & &\textbf{linear ABM}&\textbf{linear MR}&\textbf{exp ABM} &\textbf{exp MR}\\
	\hline
\textbf{PNL} &\textbf{mean}	&11.038	&14.230	&10.670	&11.081\\
 &\textbf{std dev}	&1.011	&13.653	&0.356	&0.522\\
 &\textbf{"Sharpe"}	&10.92	&1.04	&29.97	&21.23\\
 &\textbf{skewness}	&0.015	&-0.554	&0.021	&-0.987\\
 &\textbf{kurtosis}	&5.638	&4.702	&2.986	&5.941\\
 &\textbf{Jarque Bera}	&29009.6	&17181.5	&8.0	&52288.0\\
 &\textbf{VaR 5\%}	&9.435	&-10.866	&10.084	&10.168\\
 &\textbf{VaR 1\%}	&8.332	&-26.196	&9.849	&9.451\\
\hline
\textbf{Inv} &\textbf{mean}	&0.045	&335.118	&-0.013	&3.297\\
 &\textbf{std dev}	&33.253	&417.302	&7.641	&15.295\\
 &\textbf{skewness}	&0.008	&-0.626	&-0.003	&0.006\\
 &\textbf{kurtosis}	&3.004	&2.358	&3.024	&2.992\\
 &\textbf{Jarque Bera}	&1.2	&8241.5	&2.5	&0.9\\
 &\textbf{Q(T) 90\%}	&[-55,55]	&[-459,847]	&[-13,13]	&[-22,28]\\
\hline
\end{tabular}

\smallskip

Table 3A. Statistics of the PNL distribution. $\mu=1.02$, $\eta=0$.
\end{center}

\bigskip

\begin{center}
\begin{tabular}{|c|c|c|c|c|c|}
\hline
 & &\textbf{linear ABM}&\textbf{linear MR}&\textbf{exp ABM} &\textbf{exp MR}\\
	\hline
\textbf{PNL} &\textbf{mean}	&10.982	&11.581	&10.605	&10.945\\
 &\textbf{std dev}	&0.412	&1.517	&0.345	&0.446\\
  &\textbf{"Sharpe"} &26.66	&7.63	&30.74	&24.54\\
 &\textbf{skewness}	&0.031	&-2.074	&0.028	&-0.647\\
 &\textbf{kurtosis}	&3.052	&10.611	&3.011	&4.642\\
 &\textbf{Jarque Bera}	&27.2	&313076.3	&13.9	&18207.8\\
 &\textbf{VaR 5\%}	&10.308	&8.659	&10.039	&10.182\\
 &\textbf{VaR 1\%}	&10.026	&5.977	&9.807	&9.662\\
\hline
\textbf{Inv} &\textbf{mean}	&-0.003	&1.712	&0.005	&0.795\\
 &\textbf{std dev}	&5.024	&8.420	&4.574	&5.681\\
 &\textbf{skewness}	&0.001	&-0.005	&0.010	&0.008\\
 &\textbf{kurtosis}	&3.006	&2.976	&2.993	&2.980\\
 &\textbf{Jarque Bera}	&0.2	&2.8	&1.9	&2.8\\
 &\textbf{Q(T) 90\%}	&[-8,8]	&[-12,16]	&[-7,8]	&[-9,10]\\
\hline
\end{tabular}

\smallskip

Table 3B. Statistics of the PNL distribution. $\mu=1.02$, $\eta=0.0001$.
\end{center}

\bigskip

\begin{center}
\begin{tabular}{|c|c|c|c|c|c|}
\hline
 & &\textbf{linear ABM}&\textbf{linear MR}&\textbf{exp ABM} &\textbf{exp MR}\\
	\hline
\textbf{PNL} &\textbf{mean}	&10.436	&10.495	&10.000	&10.234\\
 &\textbf{std dev}	&0.335	&0.365	&0.326	&0.342\\
 &\textbf{"Sharpe"}	&31.15	&28.75	&30.67	&29.92\\
 &\textbf{skewness}	&0.038	&-0.154	&0.033	&-0.061\\
 &\textbf{kurtosis}	&3.004	&3.348	&2.990	&3.103\\
 &\textbf{Jarque Bera}	&24.0	&901.2	&18.5	&106.9\\
 &\textbf{VaR 5\%}	&9.890	&9.889	&9.468	&9.669\\
 &\textbf{VaR 1\%}	&9.666	&9.598	&9.252	&9.414\\
\hline
\textbf{Inv} &\textbf{mean}	&0.006	&0.022	&-0.006	&0.018\\
 &\textbf{std dev}	&1.678	&1.677	&1.672	&1.671\\
 &\textbf{skewness}	&0.002	&-0.001	&-0.018	&-0.001\\
 &\textbf{kurtosis}	&3.001	&3.030	&3.027	&3.023\\
 &\textbf{Jarque Bera}	&0.0	&3.8	&8.5	&2.2\\
 &\textbf{Q(T) 90\%}	&[-3,3]	&[-3,3]	&[-3,3]	&[-3,3]\\
\hline
\end{tabular}

\smallskip

Table 3C. Statistics of the PNL distribution. $\mu=1.02$, $\eta=0.001$.
\end{center}

\bigskip

In Tables 1A-3C we can observe the effect of $\eta$ on the distribution of PNL and inventory:

\begin{itemize}
\item The "Sharpe" ratio (that is mean over standard deviation) increased dramatically as $\eta$ increases for the linear utility models: the linear ABM has its "Sharpe" increased by a factor of 3 whilst the linear MR has a factor of 26. The same effect is seen for exponential MR with a factor of 1.5. However, for the exponential ABM there is not a significant change.

\item The inventory risk has an impressive reduction as $\eta$ increases: by a factor of 18 for linear ABM, 230 for linear MR, 4 for exponential ABM and 8 for exponential MR.

\item The PNL mean decreases as $\eta$ increases: -5\% for linear ABM, -25\% for linear MR, -6\% for exponential ABM and -7\% for exponential MR. 

\item None of the the PNL distributions are not normal, but in all cases the inventory distributions can be considered as Gaussian because of their low Jarque-Bera scores.
\end{itemize}

In consequence, the total effect of $\eta$ in the linear models is remarkable: it not only reduces the inventory risk, as expected, but as a welcoming side effect it also reduces the risks on the PNL distribution. Of course, nothing is free and this control on the risks comes with a reduction of the PNL mean. However, with the current framework a market-maker has enough room and tools to improve her PNL, given her risk budgets on inventory and PNL distribution.

\subsection*{The effect of $\eta$ on the linear-utility strategy}

As we have already seen, $\eta$ has some indirect control on the the PNL distribution, but we wanted to have a more detailed view of this fact. We performed 20,000 Monte-Carlo simulations for $\mu=1.00$. The rest of the parameters are as before.

\begin{center}
\includegraphics[width=6.5in, height=3.3in]{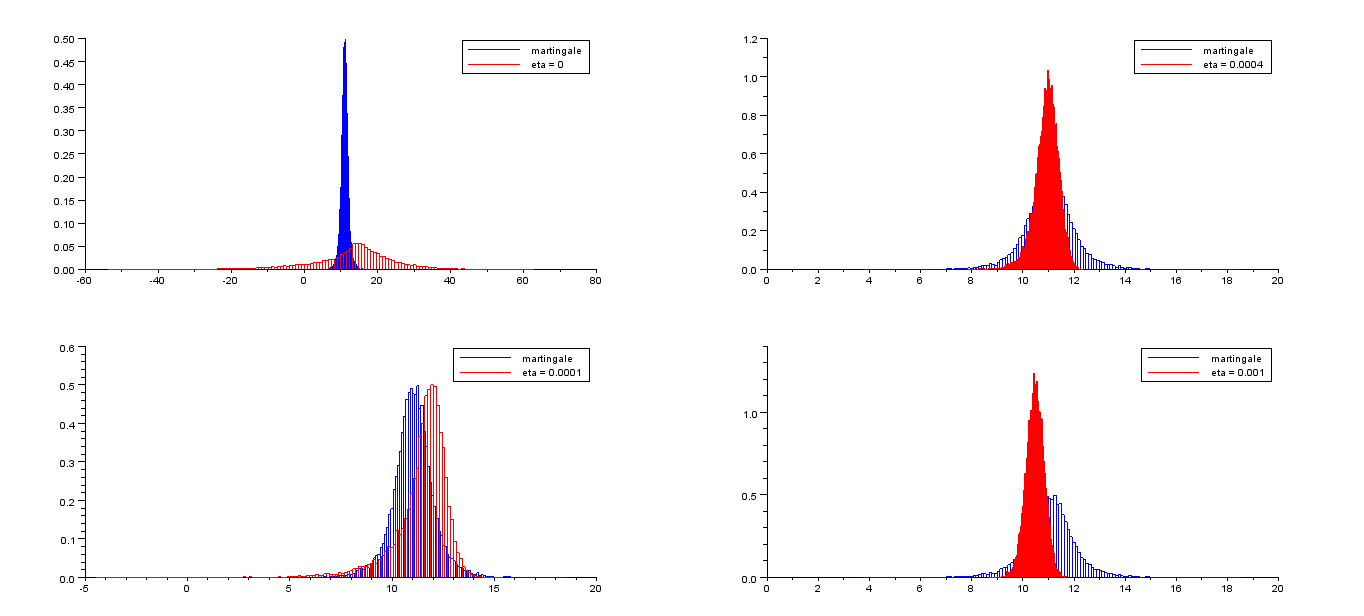}\\
Figure 6. Histogram of PNL as function of $\eta$ for linear utility function ($\mu=1.00$). Blue: Linear martingale. Red: Linear MR. Upper-Left: $\eta=0$, Lower-Left: $\eta=0.0001$, Upper-Right: $\eta=0.0004$, Lower-Right: $\eta=0.001$.
\end{center}

\begin{center}
\begin{tabular}{|c|c|c|c|c|c|}
\hline
	& Martingale	& $\eta=0$	& $\eta=0.0001$ & $\eta=0.0004$ & $\eta=0.001$ \\
\hline
\textbf{mean}	&11.035&13.386&11.452&10.930&10.474\\
\textbf{std dev} &1.002&11.800&1.320&0.473&0.356\\
\textbf{"Sharpe"} &11.01 &1.13	&8.68 &23.11 &29.42\\
\textbf{skewness} &0.009&-0.645&-2.037&-0.810&-0.131\\
\textbf{kurtosis} &5.754&5.153&10.346&5.100&3.189\\
\textbf{Jarque-Bera} &6,322.0&5,250.2&58,797.2&5,863.1&86.6\\
\textbf{VaR 5\%}	&9.447&-8.343&9.001&10.115&9.882\\
\textbf{VaR 1\%}	&8.306&-22.297&6.435&9.502&9.599\\
\textbf{Q(T) 90\%}	&[-55,56]	&[-730,732]	&[-14,14]	&[-4,4]	&[-3,3]\\
\hline
\end{tabular}

\smallskip

Table 5. Statistics of the PNL distribution for linear MR as a function of $\eta$.\\
\end{center}

As we can see in Figure 6 and Table 5, between the martingale strategy and the linear MR strategy with $\eta=0.001$ we trade 5\% of our PNL to obtain a better Sharpe of a factor of 2.7, reduce our VaR(5\%) by 8\%, our VaR(1\%) by 15\% and our inventory risk by 95\%.

\subsection*{The effect of $\gamma$ on the exponential-utility strategy}

We wanted to see if $\gamma$ can control directly the PNL distribution, as it is expected from an exponential utility function due to its variance-reduction features. In order to assess the effect of $\gamma$ separately from $\eta$, we performed 20,000 Monte-Carlo simulations for $\eta=0$ and $\mu=1.00$. The rest of the parameters are as before.

\begin{center}
\includegraphics[width=6.5in, height=3.3in]{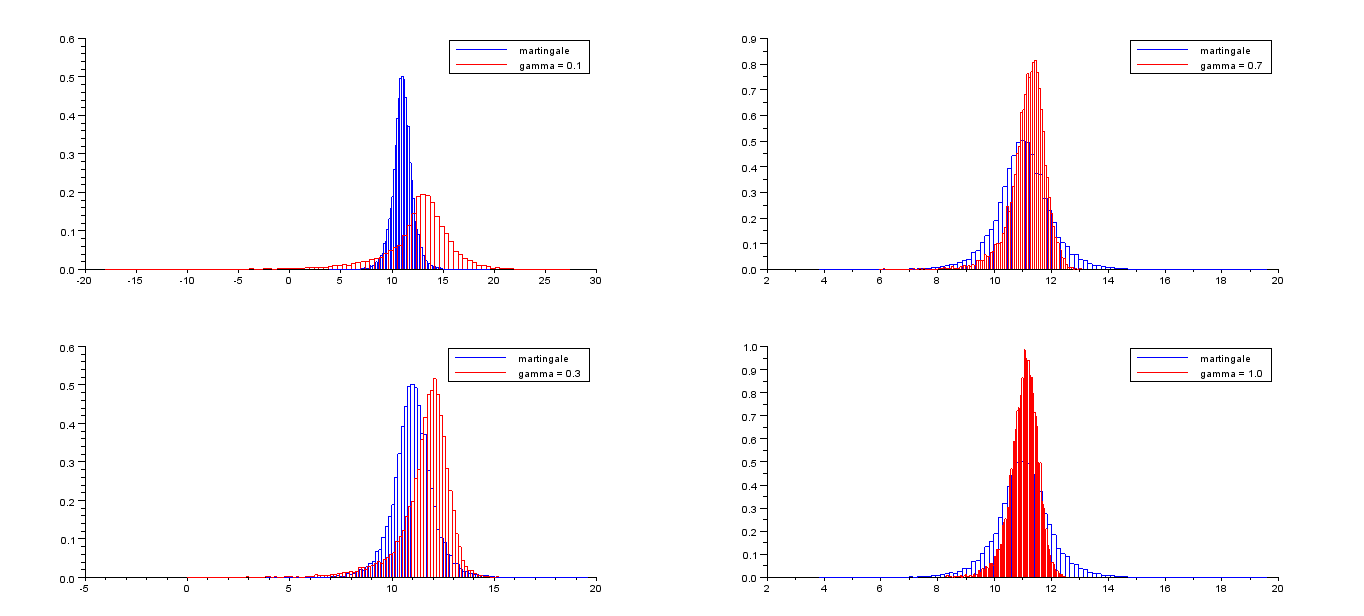}\\
Figure 7. Histogram of the PNL as a function of $\gamma$ for an exponential utility function ($\mu=1.00$).  Blue: Linear martingale. Red: Exponential MR. Upper-Left: $\gamma=0.1$, Lower-Left: $\gamma=0.3$, Upper-right: $\gamma=0.7$, Lower-Right: $\gamma=1.0$.
\end{center}

\begin{center}
\begin{tabular}{|c|c|c|c|c|c|c|c|c|c|}
\hline
	& linear martingale & $\gamma=0.1$	& $\gamma=0.3$ & $\gamma=0.7$ & $\gamma=1$\\
\hline
\textbf{mean} &11.023	&12.609	&11.581	&11.193	&11.039\\
\textbf{std dev} &0.977	&3.862	&1.230	&0.611	&0.493\\
\textbf{Sharpe}	&11.06	&3.26	&9.42	&18.32	&22.39\\
\textbf{skewness} &0.055	&-2.348	&-1.881	&-1.254	&-0.808\\
\textbf{kurtosis} &5.411	&13.242	&10.143	&6.939	&5.185\\
\textbf{Jarque-Bera} &4,853.1 &105,790.2 &54,307.3 &18,172.4 &6,154.9\\
\textbf{VaR 5}\%	&9.421	&6.225	&9.273	&10.044	&10.186\\
\textbf{VaR 1}\%	&8.330	&0.658	&7.277	&9.106	&9.572\\
\textbf{Q(T) 90\%}	&[-55,55]	&[-205,209]	&[-73,73]	&[-34,34]	&[-25,25]\\
\hline
\end{tabular}

\smallskip

Table 6. Statistics of the PNL distribution for exponential MR as a function of $\gamma$ ($\eta=0$).\\
\end{center}

As we can see in Figure 7 and Table 6, between the linear martingale strategy and the exponential MR strategy with $\gamma=1$ we have a better Sharpe of a factor of 2, reduce our VaR(5\%) by 8\%, our VaR(1\%) by 14\% and our inventory risk by 50\%. In consequence, $\gamma$ controls directly the PNL distribution and also has indirect control on the inventory, but the effect of $\eta$ seems to be stronger in both risk factors.\\

Observe that the indirect control of $\gamma$ on the inventory is less impressive than the indirect control of $\eta$ on the PNL distribution. However, with $\gamma$ the market-maker is not sacrificing any PNL at alln unlike she does with $\eta$.

\subsection*{Comparing $\eta=0.0001$ and $\gamma=0.3$}

From Figures 6-7 we have that the linear MR with $\eta=0.0001$ and the exponential MR with $\gamma=0.3$ seem to have the same (statistical) mode than the bechmark linear martingale. Here we compare these two distributions. We performed 20,000 Monte-Carlo simulations for $\mu=1.00$ and kept the other parameters unchanged.

\begin{center}
\includegraphics[width=5in, height=3in]{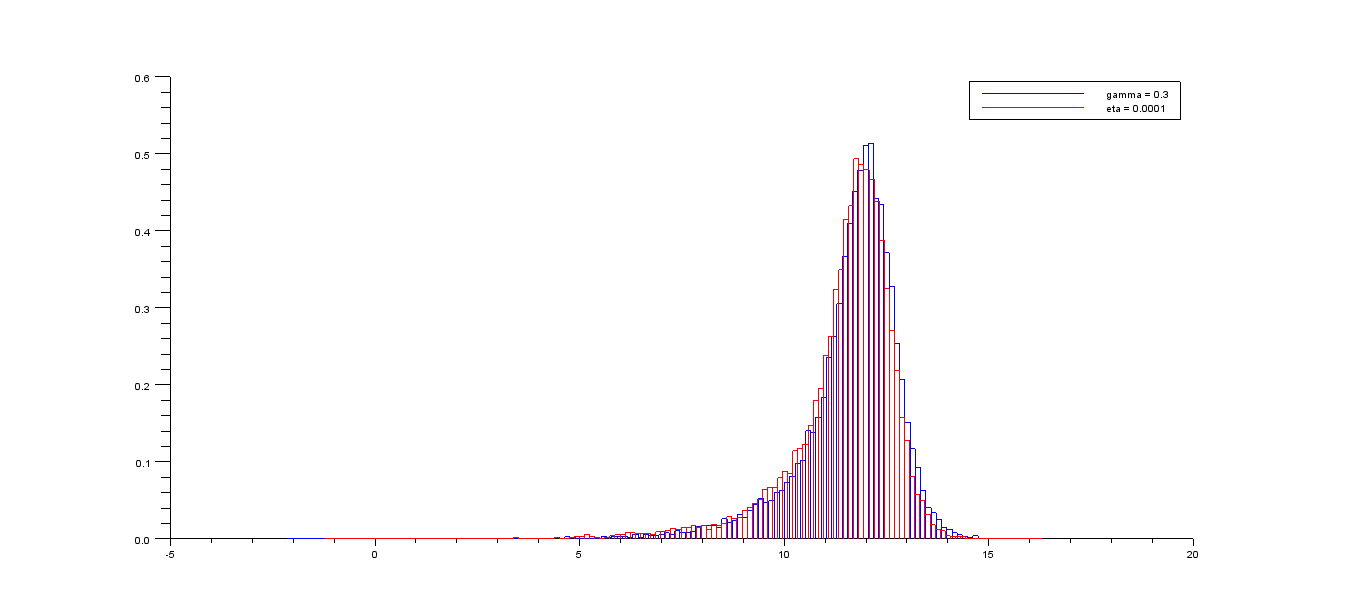}\\
Figure 8. Comparing the PNL histograms for linear MR ($\eta=0.0001$) and exponential MR ($\gamma=0.3$).\\Blue: Linear MR. Red: Exponential MR.
\end{center}

\begin{center}
\begin{tabular}{|c|c|c|c|}
\hline
	& $\gamma=0.3$	& $\eta=0.0001$ & \% change \\
\hline
\textbf{mean}&11.629&11.452&-1.53\\
\textbf{std dev}&1.235&1.320&6.91\\
\textbf{"Sharpe"}	&9.42	&8.68 &-7.86\\
\textbf{skewness}&-1.832&-2.037&11.21\\
\textbf{kurtosis}&10.165&10.346&1.78\\
\textbf{Jarque-Bera}&53967.9&58797.1&8.95\\
\textbf{VaR 5\%}&9.305&9.001&3.27\\
\textbf{VaR 1\%}&7.281&6.435&11.63\\
\textbf{Q(T) 90\%}&[-72,74]	&[-14,14]	&-80.8\\

\hline
\end{tabular}

\smallskip

Table 7. Statistical comparison between linear MR $\eta=0.001$ and exponential $\gamma=0.3$. The \% change is the variation of the linear MR with respect to the exponential MR. Skewness change is in absolute value.
\end{center}

As we can see in Table 7, seems that the exponential MR has better control on the PNL distribution than the linear MR: higher mean, "Sharpe" and quantiles at 1\% and 5\% (VaR); smaller standard deviation, (absolute) skewness and kurtosis. However, the inventory of exponential MR is much higher than the inventory of linear MR (5 times bigger).

\section{Conclusions}

\subsection*{On the market-making model}

\begin{itemize}
\item In Theorem \ref{thm-exp} we generalised the Avellaneda and Stoikov \cite{ave} approach for an exponential utility to any Markov process, provided its conditional expectation $\E_{t,s}[S(T)]$ is affine in $s$, which includes processes like arithmetic Brownian motion with drift or Ornstein-Uhlenbeck process. This allowed us to assess the effect of directional bets on the market mid-price on the PNL distribution of a high-frequency market-maker. Moreover, we also showed that the results of Lehalle \emph{et al} \cite{lehalle} can be generalised to arithmetic Brownian motions with time-dependent drift and volatility, and that their approach cannot be extended further with the ansatz they chose.

\item If instead of an exponential utility function we choose a linear utility function with inventory constraints, in Theorems \ref{thm-lin}-\ref{thm-pi} we showed that using the same stochastic-control approach we can find the optimal controls (i.e. the market-maker bid and ask quotes) for very general Markov processes, even with jumps (e.g. L\'evy processes), provided the inventory-risk penalty $\pi(s)$ is chosen appropriately to ensure the boundedness of the value function $u(t,s,q,x)$. Moreover, the optimal controls in the linear case are independent of the volatility of the asset, which is very hard to estimate for high-frequency data, and as such the linear case is easier to calibrate with real data than the exponential case.

%\item In our numerical simulations we compared the martingale case to an Ornstein-Uhlenbeck process, which is a mean-reverting process. However, our approach could be considered as \emph{almost model-free}: regardless on how does the market-maker compute the value $\E_{t,s}[S(T)]$, once this quantity is known for all times she can plug it into our model, which would in turn compute the optimal quotes that optimise her PNL under inventory constraints.

\item Our approach, although based on optimal-control and nonlinear-PDE techniques, is very intuitive: choose the right \emph{ansatz} for the solution, compute the (implicit) controls, plug them into the equation, separate the equation into smaller and easier parts and solve them all to have the explicit form of the solution and the controls.  Moreover, if the full equation is not explicitly solvable, approximate the jump part due to inventory: the resulting control is thus optimal for a sub-solution, and as such we are controlling the PNL utility function from below, which translates into optimal quotes to reduce losses.

\end{itemize}

\subsection*{On the role of the parameters $\eta$ and $\gamma$}

\begin{itemize}
\item If $\eta$ increases then the inventory risk decreases and eventually becomes negligeable. This was expected because the parameter $\eta$ was added as an inventory penalty. However, as a nice side effect we have that $\eta$ also reduces the risk on the PNL, in the sense that the variance and the kurtosis of the PNL distribution decrease as $\eta$ increases. In other words, $\eta$ controls perfectly the inventory risk, and by doing so it indirectly controls the PNL distribution.

\item For $\gamma$ we have the a similar effect, but the other way around. As $\gamma$ increases the risk on the PNL distribution decrease, in particular the first four moments and the VaR. Moreover, as a side effect the inventory risk decreases as well. This implies that $\gamma$ directly controls the PNL distribution and indirectly controls the inventory risk.

\item This mirror-like, intertwined role of $\eta$ and $\gamma$ highlights a intimate relation between inventory risk and the risk on the PNL distribution: if one decreases then necessarily the other has to do the same. This implies that a market-maker who chooses to reduce the risk on her PNL distribution necessarily reduces her inventory risk; conversely, a market-maker who reduces her inventory risk also reduces the risk on her PNL distribution. However, this risk reduction also implies a reduction on the average PNL.

\item It seems that the control provided by $\eta$ on the linear MR is better than the control provided by $\gamma$ on the exponential MR. However, this claim has to be taken with a grain of salt because, for $\eta=0.0001$ and $\gamma=0.3$ (where qualitatively both distributions are comparable to the martingale benchmark) the linear MR is less risky in terms of inventory whilst the exponential MR is less risky in terms of PNL distribution. That said, our numerical simulations show that an increase on $\eta$ (from 0.0001 to 0.001) renders a more drastic reduction on inventory risk and stronger structural changes on PNL distribution than an increase on $\gamma$ of the same order of magnitude (from 0.1 to 1).

\end{itemize}

\subsection*{On the directional bets and the risk profiles}

As we have shown in our simulations, there is a clear (and expected) relation between risk and reward. 

\begin{itemize}
\item When the market-maker makes a directional bet she improves her PNL up to 25\% with respect to the martingale benchmark. However, by doing so she has to accept more risk, either on her PNL distribution (measured in terms of variance, skewness and kurtosis) or on her inventory.

\item After a directional bet, the market maker can trade some of her excess PNL (over the martingale benchmark) for a direct control on her risks. If she chooses to gain direct control on her PNL distribution directly (resp. inventory risk) then she gains some reduction on her inventory risk (resp. PNL distribution), but the latter cannot be controlled directly. In that spirit, the market-maker can choose to use an exponential MR strategy if she prices more dearly her risk on the PNL distribution and a linear MR strategy if her biggest concern is the inventory risk.

%\item The inventory-risk-aversion parameter $\eta$ seems a better choice for global risk control (i.e. the joint effect of PNL distribution and inventory risk) than $\gamma$. Indeed, the side effect of $\eta$ on the PNL distribution seems to be more important than the side effect of $\gamma$ on the inventory risk: compare e.g. $\eta=0.0007$ on Table 5 vs $\gamma=1$ on Table 6.

\item In summary, directional bets enhance the PNL of the market-maker but add extra risk on her inventory and PNL distribution; both risks are positively correlated, so if she reduces one directly the other is indirectly reduced as well. Therefore, a market-maker in practice has to assess three factors, i.e. PNL increase, inventory risk and PNL distribution, in order to choose her optimal trading strategy because, in the current set-up at least, there is no strategy that is optimal in all three factors.

\end{itemize}

\subsection*{Further developments}

\begin{itemize}
\item In our model there are no market orders, just limit orders. For the simulations we assumed that if $\delta^+_\ast\le 0$ (resp. $\delta^-_\ast\le 0$) then the market-maker sends a selling (resp. buying) market order that is executed at the mid-price, i.e. we assumed that the market spread is zero and their impact (or cost) on the PNL is $\vert \delta^\pm\vert\ge0$.

\item For a detailed market-impact analysis, we should add the market spread as another state variable and consider that the market-maker's limit orders affect both the market spread and the mid-price. For example, if she improves the best ask (resp. bid) then she reduces the market spread by one tick and pushes down (resp. up) the mid-price by half a tick.

\item Another possibility is to incorporate market orders directly into the model and making a clear distinction between market and limit orders (see e.g. Guilbaud and Pham \cite{pham-hft}). However, the current framework does not seem to be easily extended for that purpose, which suggests that a new framework is needed. We are currently working on that direction.

\item We have assumed that the mid-price is continuous, which can be interpreted in the discretisation for our simulations as assuming that the bid-ask spread of the market-maker is large with respect to small changes on the mid-price. This is true for assets whose spread is large with respect to the tick size (e.g. equities), but not for futures for which the spread is in average 1-2 ticks, so the smallest price move can make the market-maker's quotes cross the spread. Therefore, another framework is necessary to deal with assets whose spread and tick size are comparable. We will address this problem in a future work.

\item Our mid-price $S(t)$ is a Markov processes, just like in Avellaneda and Stoikov \cite{ave} and Lehalle \emph{et al}. This means that the infinitesimal generator is local, $S(t)$ has zero auto-correlation and the Hamilton-Jacobi-Bellman equation is a nonlinear PDE. Now, if we consider a mid-price process with non-zero autocorrelation (e.g. fractional Brownian Motion, multifractal processes or Hawkes process) then its infinitesimal operator is non-local and its HJB equation is an integro-differential equation. However, in this case we cannot invoke the Feynman-Kac formula, which means that our approach is no longer valid and another model is necessary (see e.g. Cartea and Jaimungai \cite{cartea-bs} and \cite{cartea-risk}).

\end{itemize}

\subsection*{Acknowledgements}

The authors would like to thank Prof. Huy\^en Pham (University Paris-Diderot, France) and Prof. Mathieu Rosenbaum (University Pierre et Marie Curie, France) for their suggestions and kind advice.

\end{document}